\documentclass[natbib]{svjour3}             

\smartqed  
\usepackage{graphicx,amsmath,amssymb}
\usepackage{graphicx}
\usepackage{aps-bibstyle}
\usepackage{epsfig}
\usepackage{color}
\usepackage{epstopdf}
\newcommand{\apj}{{Astrophys. J.}}

\newcommand{\solphys}{{Solar Phys.}}
\begin{document}

\title{Nonlinear effects in resonant layers in solar and space plasmas}


\author{Istvan Ballai         \and
        Michael S. Ruderman 
}

\authorrunning{Ballai and Ruderman}


\institute{I. Ballai \and M.S. Ruderman \at
              Solar Physics and Space Plasmas Research Centre, Dept. 
	      of Applied Mathematics, University of Sheffield, Hounsfield 
	      Road, Hicks Building, Sheffield, S3 7RH  \\
              Tel.: +44-114-22-23833/23717\\
              \email{i.ballai@sheffield.ac.uk}           
}

\date{Received: date / Accepted: date}

\maketitle

\begin{abstract}
The present paper reviews recent advances in the theory of
nonlinear driven magnetohydrodynamic (MHD) waves in slow and
Alfv\'en resonant layers. Simple estimations show that in the
vicinity of resonant positions the amplitude of variables can grow
over the threshold where linear descriptions are valid. Using the
method of matched asymptotic expansions, governing equations of
dynamics inside the dissipative layer and jump conditions across the
dissipative layers are derived. These relations are essential when
studying the efficiency of resonant absorption. Nonlinearity in
dissipative layers can generate new effects, such as mean flows,
which can have serious implications on the stability and efficiency
of the resonance.
\end{abstract}

\section{Introduction}
\label{sec:1}

The dynamical response of the plasma in the solar atmosphere to 
rapid changes can be manifested through wave propagation along and 
across the magnetic field. Many of these waves are in the
magnetohydrodynamic (MHD) threshold (waves with periods larger than the ion collisional time and wavelengths larger than the mean free path of ions) and they are observable by the new
generation of space- and ground-based telescopes in almost all
regions of the solar atmosphere \cite[for extensive reviews on wave
observation see, e.g.,][]{acton1981,nak05,banerjee07}.

One of the most fundamental characteristics of solar and space
plasmas is their very high degree of inhomogeneity along and across
magnetic fields. It is well known that the properties of MHD waves 
are strongly modified by the plasma inhomogeneity. In particular,
when the inhomogeneity of the medium is transversal to the direction 
of wave propagation a new phenomenon, called 
{\it resonant absorption}, can appear.

While homogenous plasmas have a spectrum of linear eigenmodes which
can be divided into slow, fast and Alfv\'en subspectra, with the
slow and fast subspectra having discrete eigenmodes and the
subspectrum of Alfv\'en waves being infinitely degenerated, in
inhomogeneous plasmas the three subspectra are substantially changed. {\bf According to spectral theories of waves in inhomogeneous plasmas
\cite[see, e.g.,][]{goed83,goed04} the spectra of slow and Alfv\'en waves become continuous while the spectrum corresponding to fast waves becomes discrete.} 
Eigenfunctions that correspond to 
frequencies in the continuum spectra are improper as they contain a 
non-square integrable singularity at the resonant position.

According to the accepted {\bf resonant wave theories}, effective energy transfer
between an energy carrying wave and the plasma occurs if the
frequency of the wave falls into the slow or
Alfv\'en continuum, i.e. at the slow or Alfv\'en resonances.

The process of resonant interaction of waves and the possibility of
energy transfer between global and local waves has been studied
intensively for the past few decades. First resonant absorption was
studied as a means of supplementary heating fusion plasmas, but was
later rejected due to technical difficulties 
\cite[see, e.g.,][]{tataronis73,chen74,poedts89,vaneester91}.

In the Earth's magnetosphere resonant MHD wave coupling is believed
to generate low frequency pulsation or energize ULF waves 
\cite[see, e.g.][]{lanzerotti73,southwood74,southwood83,rud98,erdelyi03,
taroyan02,taroyan03a,taroyan03b}. 
Within the context of solar physics
Ionson (1978) was the first who proposed resonant absorption as a
possible mechanism for coronal heating. His idea was further
developed in the last few decades by many authors 
\cite[see, e.g.][ etc.]{kuperus81,davila87,hollweg91,goossens91, 
ofman95,erdelyi94,erdelyi1995, erdelyi96b,erdelyi1997,erdelyi1998,ballai00a}, however it became clear soon that resonant
absorption alone cannot explain the very high temperature of the
corona. Resonant absorption was also used to explain the loss of
power in p-modes when interacting with sunspots in the solar
photosphere 
\cite[see, e.g.,][]{hollweg88,lou90,goossens92,spruit92,erdelyi94,keppens94,erdelyi1996a,tirry98a,tirry98b}.

Recently resonant absorption acquired a new application in {\it
coronal seismology} where the rapid damping (with damping times of
the order of a few periods) of kink oscillations in coronal loops is
explained in terms of resonant interaction of global kink modes and
the local Alfv\'en waves \cite[][]{ruderman02,goossens02,goossens06,dym06,erd+ver07,verth08,terr10}. 
Coronal seismology is dependent on theoretical relations
(dispersion relations) which link plasma parameters, such as the
plasma density, to wave parameters, such as the wave frequency, in a
precise way. Generally, plasma parameters are determined from wave
parameters, which themselves are determined observationally. The
dispersion relations for many plasma structures under
the assumptions of ideal MHD are well known;
they were derived long before accurate EUV observations were
available \cite[e.g.][]{edw83,rob84} using
simplified models within the framework of ideal and linear
MHD. Although the realistic interpretation of
many observations are made difficult by the spatial and temporal
resolution of present satellites not being adequate, considerable
amount of information about the state of the plasma and the
structure and magnitude of the coronal magnetic field have been
already obtained.

At resonance, amplitudes of oscillations (i.e. the energy density)
can grow without limits, however even a small amount of dissipation
is able to prevent the unlimited increase of the amplitude. The
presence of dissipation causes transforming the wave 
energy into heat. On the other hand solar and space
plasmas are media with very high Reynolds numbers (i.e. weakly
dissipative media), which means that the dissipation is an important
ingredient in the description of dynamics only in a narrow region
around the resonant magnetic surfaces called {\it dissipative
layer}. Outside this layer the dynamics is described by the ideal
MHD. This natural structuring of the domain of interest allows us to
use the method of matched asymptotic expansions 
\cite[][]{nayfeh81,bender91} to describe the
problem of resonant absorption of MHD waves. The essence of this
method resides in solving ideal MHD equations outside the
dissipative layer, the non-ideal equations inside the dissipative layer 
(which contains the resonant surface), and matching
the two sets of solutions in overlapping regions (for details see the review by Goossens et al. 2010, this issue).

The dissipation of energies in the solar atmosphere is a delicate problem as the
dominance of a certain process over all other possible ones depends
on the region where the dynamics will be described and also on the
very physics of the problem under consideration. That is why the
dynamics of waves in the solar atmosphere is going to be influenced
by different dissipation mechanisms in, e.g. the photosphere and solar
corona. A full analysis of particular dissipative mechanisms
used in the present paper will be given in the next section.

Nonlinearity was systematically overlooked in many 
studies related to wave dynamics because the mathematical treatment 
of nonlinear problems is cumbersome, and because the 
complicated mathematical analysis can often obscure the physics 
and, therefore, make the results unusable for observers. Dynamics 
occurring over extended scales (as in the solar atmosphere) are likely to 
encounter situations which can lead to the increase in amplitudes resulting in 
breakdown of linear description. Very often the wave amplitudes can 
increase due to the change of the medium properties, e.g. 
acoustic waves propagating from the photosphere to the corona can steepen 
into shocks due to the decrease of the density with height.

A great advantage in tackling the complicated problem of resonant
absorption is the notion of connection formulae introduced for the
first time by \cite{sak91}. The accuracy of the method by numerics was addressed by \cite{stenuit95}. 
This approach assumes that the thin
dissipative layer is treated as a surface of discontinuity with the 
dynamics at both sides of the discontinuity fully described by the 
ideal MHD. The system of non-ideal MHD equations describing the 
dynamics {\it inside} the dissipative layer is used to obtain the 
connection formulae that serve as boundary conditions at the surface 
of discontinuity in the same way as the Rankine-Hugoniot jump conditions 
for shocks.

{\bf A great deal of understanding of the process of resonant absorption came from recent numerical investigations \cite[see, e.g.,][]{ofm94,ofman95,ofm96,poed97}, where issues like higher dimension resonance,
time evolution of the resonance or randomly driven resonance were discussed in great detail.}

The paper is organized as follows. In Section.~\ref{sec:2} we introduce
the equations used to describe the dynamics of waves and present the
mathematical and physical tools employed in the mathematical
presentation of the problem. In Section 3 we introduce scalings and dimensionless quantities required to quantify the importance 
of nonlinearity. The problem of nonlinear slow resonant
waves is presented in Section 4, where we describe the nonlinear character of waves near resonance and we provide the jump conditions near the resonance. In Section 5
we show that there are two particular cases where the jump conditions can be simplified into an explicit form. The problem of nonlinear resonant Alfv\'en
waves is discussed in Section 6. As an application to the nonlinear theory reviewed in this paper we will apply the equations to study the efficiency of resonance in the limiting cases of weak and strong nonlinearity. 
A direct consequence of the
nonlinear framework used in our analysis is the generation of mean shear
flows at both resonances, which will be presented in Section 8.

\section{Basic equations}
\label{sec:2}

Solar and space plasmas are far from being an ideal environment with
the plasma dynamics being affected by many dissipative and
dispersive effects. As stressed earlier, the right choice of 
dissipative mechanism depends on the location where a particular
physical phenomenon takes place as well as on the nature of
this phenomenon. For the present paper, we are going to consider viscosity, 
electrical resistivity and thermal conduction. Dissipative processes are 
weak in the solar atmosphere; this means that the diffusion coefficients 
are small. The rate of dissipation, however, is dependent on the local
spatial scales.

Many phenomena (e.g. resonant absorption, current sheets or turbulences)
are inherently non-ideal and very often nonlinear as they are strongly
influenced by dissipative and/or dispersive effects. In particular,
dissipation is important for nonlinear dynamical processes because
large-scale motions rapidly lead to the formation of small-scale 
structures, which corresponds to singularities in the ideal theory.

Viscosity and thermal conductivity are linked to hydrodynamical
processes, while electrical conductivity and Hall dispersion are
related to the presence of the electric and magnetic 
fields. In the context of solar physics, the general effect of 
dissipation and dispersion is to relax the accumulation of wave energy 
in a system. The relaxation can be performed by, e.g. converting 
the wave energy at resonance into heat by viscous or resistive 
dissipation or thermal conduction (via resonant absorption), or the 
dispersion of energy over a larger area by the Hall effect. 
The relaxation caused by dissipation and dispersion prevents the formation 
of singularities (entities abhorred by nature).

The key quantities in our discussion are the dimensionless products
$\omega_{i}\tau_{i}$ and $\omega_{e}\tau_{e}$, where $\omega_{i(e)}$
is the ion (electron) cyclotron frequency and $\tau_{i(e)}$ is the
mean ion (electron) collision time. Since viscosity is mainly due to
ions, the product $\omega_{i}\tau_{i}$ is important in describing 
viscosity. On the other hand, thermal and electrical conductivity are 
effects attributed mainly to electrons, so the product $\omega_{e}\tau_{e}$ 
appears in the description of these two dissipative mechanisms.

The classical \cite{bra65} expression for the viscosity tensor reads
\begin{equation}
\boldsymbol{\mathsf{S}}=\sum_{i=0}^4 \eta_i\boldsymbol{\mathsf{S}}_i,
\label{eq:2.1}
\end{equation}
where the coefficients $\eta_0$, $\eta_1$ and $\eta_2$ describe viscous 
damping and $\eta_3$ and $\eta_4$ are non-dissipative coefficients related 
to wave dispersion due to the finite ion gyroradius. The quantities 
$\boldsymbol{\mathsf{S}}_i$ are given in terms of the unit vector along the 
equilibrium magnetic field and the velocity vector. The viscous force in 
plasmas is given by $\nabla\cdot\boldsymbol{\mathsf{S}}$\/. In general the
expressions for $\eta_i$ ($i = 1,\dots,4$) and $\boldsymbol{\mathsf{S}}_i$ are 
quite complicated and we do not write them down. The coefficient of 
compressional viscosity $\eta_0$ is given by
\begin{equation}
\eta_0\approx\frac{\rho k_BT\tau_i}{m_p}, 
\label{eq:2.2}
\end{equation}
where $\rho$ and $T$ are the density and temperature of the plasma, $k_B$ is 
the Boltzmann constant and $m_p$ is the proton mass. When $\omega_i\tau_i\ll 1$ 
as in the solar photosphere, the other coefficients in Eq.~(\ref{eq:2.1}) are 
given by
\begin{equation}
\eta_1 \approx \eta_2 \approx \eta_0, \quad
\eta_3 \approx 2\eta_4 \approx 0.8\eta_0(\omega_i\tau_i). 
\label{eq:2.2a}
\end{equation}
It can be seen that $\eta_3,\eta_4 \ll \eta_0$\/, so that the last two terms in
Eq.~(\ref{eq:2.1}) can be neglected. Then the viscous force is given by the
approximate expression
\begin{equation}
\vec{F}_{\rm vis} = \nabla\cdot\boldsymbol{\mathsf{S}} \approx
\eta_0\big(\nabla^2\vec{v} + \frac{_1}{^3}\nabla\nabla\cdot\vec{v}\big),
\label{eq:2.2b}
\end{equation}
where $\vec{v}$ is the plasma velocity. This expression for the viscous force 
is isotropic in the sense that the force is independent of the magnetic field 
direction.

In the solar chromosphere and, especially in the corona, the condition 
$\omega_i\tau_i\gg 1$ is satisfied.  In this case the remaining four 
dissipative coefficients are given by
\begin{equation}
\eta_1=\frac{\eta_0}{4(\omega_i\tau_i)^2}\/, \quad \eta_2=4\eta_1\/, \quad
\eta_3 = \frac{\eta_0}{2\omega_i\tau_i}\/, \quad \eta_4 = 2\eta_3 .
\label{eq:2.3}
\end{equation}
Now $\eta_1,\eta_2,\eta_3,\eta_4 \ll \eta_0$\/, so that, in 
particular, the compressional viscosity strongly dominates the shear viscosity. 
As a result all terms in Eq.~(\ref{eq:2.1}) can be neglected in comparison with 
the first one, and we arrive at the following approximate expression for the
viscous force, 
\begin{equation}
\vec{F}_{\rm vis} = \nabla\cdot\boldsymbol{\mathsf{S}} \approx
\eta_0\nabla\cdot\left(\vec{b}_0\otimes\vec{b}_0 - 
\frac{_1}{^3}\boldsymbol{\mathsf{I}}\right)
[3\vec{b}_0\cdot(\vec{b}_0\cdot\nabla)\vec{v} - \nabla\cdot\vec{v}].
\label{eq:2.3a}
\end{equation}
Here $\vec{b}_0$ is the unit vector in the magnetic field direction, 
$\boldsymbol{\mathsf{I}}$ is the unit tensor, and $\vec{f} \otimes\vec{g}$
denotes the dyadic product of vectors $\vec{f}$ and $\vec{g}$\/.

However, Eq.~(\ref{eq:2.3a}) can be used only in slow dissipative layers. 
The compressional viscosity is identically zero for Alfv\'en waves. As a 
result, it cannot remove the ideal Alfv\'en singularity 
\cite[][]{ofm94,erd95,ruderman96,mocanu08}. When studying the motion in 
Alfv\'en dissipative layers we have to keep the second and third terms in 
Eq.~(\ref{eq:2.1}). On the other hand, in spite that 
$\eta_1,\eta_2 \ll \eta_0$\/, we can neglect the first term in 
Eq.~(\ref{eq:2.1}). As for the fourth and fifth terms, they can be neglected
because they do not provide dissipation. In addition, we can neglect the 
spatial variation of the unit vector in the magnetic field direction,
$\vec{b}$\/, in the Alfv\'en dissipative layer. As a result, we obtain the 
following expression for the viscous force, 
\begin{eqnarray}
\vec{F}_{\rm vis} &\approx& \eta_1\nabla\cdot(\boldsymbol{\mathsf{S}}_1
   + 4\boldsymbol{\mathsf{S}}_2) \approx \eta_1\{\nabla^2\vec{v}
   + 3\vec{b}_0\nabla^2(\vec{b}_0\cdot\vec{v}) 
   + 4\vec{b}_0\/\/\vec{b}_0\cdot\nabla\nabla\cdot\vec{v} \nonumber\\
&+& 3(\vec{b}_0\cdot\nabla)[(\vec{b}_0\cdot3\nabla) 
   + \nabla(\vec{b}_0\cdot\vec{v})]
   + \nabla[\vec{b}_0\cdot\nabla(\vec{b}_0\cdot\vec{v})] \nonumber\\
&-& 15\vec{b}_0\/\/\vec{b}_0\cdot
    \nabla[(\vec{b}_0\cdot\nabla(\vec{b}_0\cdot\vec{v})]\},
\label{eq:2.3b}
\end{eqnarray}
which is still rather complicated. However,  in 
Alfv\'en dissipative layers $\nabla\cdot\vec{v}\approx 0$ which enables us to neglect the third term
on the right-hand side of Eq.~(\ref{eq:2.3b}). In addition, in Alfv\'en
dissipative layers large gradients can develop only in the direction 
perpendicular to the equilibrium magnetic field. This observation enables us 
to neglect also the fourth, fifth and sixth terms on the right-hand side of 
Eq.~(\ref{eq:2.3b}). Finally, the dominant component of the velocity in 
Alfv\'en dissipative layers is the component orthogonal to the equilibrium 
magnetic field, while the component along the magnetic field remains small. 
This means that we can also neglect the second term on the right-hand side of 
Eq.~(\ref{eq:2.3b}). As a result the very simple expression for the 
viscous force that can be used in Alfv\'en dissipative layers is
\begin{equation}
\vec{F}_{\rm vis} \approx \eta_1\nabla^2\vec{v}.
\label{eq:2.3c}
\end{equation}

The wave energy can be also dissipated due to thermal conduction. In what
follows we assume that the ions end electrons have equal temperatures. In this case the
general expression for the heat flux $\vec{q}$ is given by
\begin{equation}
\vec{q} = -\kappa_\parallel\nabla_\parallel T - \kappa_\perp\nabla_\perp T
    - \kappa_\wedge\vec{b}_0\times\nabla T ,
\label{eq:2.3d}
\end{equation}
where $\kappa_\parallel$\/, $\kappa_\perp$ and $\kappa_\wedge$ are the 
parallel, perpendicular and skew coefficients of thermal conduction,
$\nabla_\parallel = \vec{b}_0(\vec{b}_0\cdot\nabla)$\/,
$\nabla_\perp = \nabla - \vec{b}_0(\vec{b}_0\cdot\nabla)$, and $T$ is the 
plasma temperature related to the plasma density, $\rho$\/, and pressure, 
$p$\/, by the ideal gas law,
\begin{equation}
p = \frac{k_B}{m_p}\rho T .
\label{eq:2.3e}
\end{equation}
For the electron-proton plasmas, the expression for $\kappa_\parallel$ is given by
\begin{equation}
\kappa_\parallel \approx \frac{3k_B^2 \rho T\tau_e}{m_p m_e},
\label{eq:2.3f}
\end{equation}
where $m_e$ is the electron mass. When $\omega_e\tau_e \ll 1$ as in the lower
part of the solar photosphere, $\kappa_\perp \approx \kappa_\parallel$ and 
$\kappa_\wedge \sim (\omega_e\tau_e)\kappa_\parallel \ll \kappa_\parallel$\/.
As a result the approximate expression for the heat flux reads
\begin{equation}
\vec{q} = -\kappa_\parallel\nabla T .
\label{eq:2.3g}
\end{equation}
This expression is isotropic in the sense that it is independent of the 
magnetic field direction.

When $\omega_e\tau_e \gg 1$ as in the solar chromosphere and corona, 
$\kappa_\perp\sim (\omega_e\tau_e)^{-2}\kappa_\parallel\ll \kappa_\parallel$\/,
$\kappa_\wedge\sim (\omega_e\tau_e)^{-1}\kappa_\parallel\ll \kappa_\parallel$\/,
and the first term in Eq.~(\ref{eq:2.3d}) strongly dominates over two other
terms. Therefore we can use the approximate expression for the heat flux
\begin{equation}
\vec{q} = -\kappa_\parallel\nabla_\parallel T .
\label{eq:2.3h}
\end{equation}
We now see that the heat flux is in the magnetic field direction, and magnetic
surfaces act as thermal insulators.

We write the expression of the generalized Ohm's equation in the form
\begin{equation}
\vec{E} + \vec{v}\times\vec{B} = \frac{\vec{j}}{\sigma} +
   \frac{m_p}{\rho e}\vec{j}\times\vec{B},
\label{eq:2.3i}
\end{equation}
where $e$ is the elementary electrical charge, $\sigma$ the electrical
conductivity, $\vec{j}$ the electrical current density, and $\vec{E}$ and 
$\vec{B}$ are the electric and magnetic field, respectively. The first term on
the right-hand side of Eq.~(\ref{eq:2.3i}) is related to the plasma
resistivity. It disappears in the limit of infinitely conducting plasmas where
$\sigma \to \infty$\/. The second term on the right-hand side of 
Eq.~(\ref{eq:2.3i}) describes the Hall current related to the account of the
ion inertia. A more general form of the Omh's equation contains also the term
proportional to $\vec{B}\times(\vec{j}\times\vec{B})$ related to the so-called
Cowling conductivity (see, e.g. Priest 2000). This term is identically 
zero for a fully ionized plasma, while in partially ionized plasmas, can be 
important only when the electron gyrofrequency is large and the plasma is 
sufficiently rarified, so that the mean collision time between electrons and 
neutrals is large. On this ground we disregard this term. In what follows we
neglect the displacement current in Maxwell's equations and use Ampere's law
to relate the electrical current density and magnetic field,
\begin{equation}
\mu_0\vec{j} = \nabla\times\vec{B},
\label{eq:2.3j}
\end{equation}
where $\mu_0$ is permeability of free space.

In our analysis we use the system of MHD equations
\begin{subequations}
\label{eq:2.4}
\begin{equation}
\frac{\partial\rho}{\partial t} + \nabla\cdot{\rho\vec{v}}=0,
\label{eq:2.4a}
\end{equation}
\begin{equation}
\rho\left(\frac{\partial\vec{v}}{\partial t} + \vec{v}\cdot\nabla\vec{v}\right)
   = -\nabla p+\frac{1}{\mu_0}(\nabla\times\vec{B})\times\vec{B} 
   + \vec{F}_{\rm vis}, 
\label{eq:2.4b}
\end{equation}
\begin{equation}
\frac{\partial\vec{B}}{\partial t} = -\nabla\times\vec{E},
\label{eq:2.4c}
\end{equation}
\begin{equation}
\frac{\partial}{\partial t}\left(\frac{p}{\rho^{\gamma}}\right) +
\vec{v}\cdot\nabla\left(\frac{p}{\rho^{\gamma}}\right) =
-\frac{\gamma-1}{\rho^{\gamma}}{\cal L}.
\label{eq:2.4d}
\end{equation}
\end{subequations}
In these equations $\vec{F}_{\rm vis}$ is given either by Eq.~(\ref{eq:2.2b}), Eq.~(\ref{eq:2.3a}) or Eq.~(\ref{eq:2.3c}), and $\vec{E}$ is given by
Eq.~(\ref{eq:2.3i}); $\gamma$ is the ratio of specific heats or adiabatic
index, and ${\cal L}$ is the loss function given by
\begin{equation}
{\cal L} = \nabla\cdot\vec{v} - \frac{\vec{j}^2}\sigma -
   \boldsymbol{\mathsf{S}}:\nabla\vec{v},
\label{eq:2.5}
\end{equation}
where $\boldsymbol{\mathsf{S}}$ is the viscosity tensor and the colon 
indicates the double summation,
\begin{equation}
\boldsymbol{\mathsf{S}}:\nabla\vec{v} = 
   S_{ij}\frac{\partial v_i}{\partial x_j}.
\label{eq:2.6}
\end{equation}
Recall that the magnetic field is solenoidal, i.e.
\begin{equation}
\nabla\cdot\vec{B} = 0.
\label{eq:2.7}
\end{equation}
This equation can be considered as an initial condition imposed on the magnetic
field because, if it is satisfied at $t = 0$, then it follows from 
Eq.~(\ref{eq:2.4c}) that it is satisfied for any $t > 0$. If we neglect the
Hall term in Eq.~(\ref{eq:2.3i}) then the induction equation~(\ref{eq:2.4c})
reduces to
\begin{equation}
\frac{\partial\vec{B}}{\partial t} = \nabla\times(\vec{v}\times\vec{B})
   + \eta\nabla^2\vec{B},
\label{eq:2.8}
\end{equation}
where $\eta = 1/\mu_0\sigma$ is the coefficient of magnetic diffusion. {\bf These equations must be supplemented by the Ohm's law (\ref{eq:2.3i}) expressing the connection between the electric field and magnetic field.}

\section{General discussion and dimensionless parameters}
\label{sec:3}

Traditionally the importance of dissipative processes in MHD is
characterized by dimensionless parameters calculated as ratio of
corresponding dissipative terms to dynamical terms.
The importance of viscosity is quantified by the viscous Reynolds number. 
In the previous section we have seen that the coefficients of compressional and
shear viscosity can be quite different, so that we have to introduce two
Reynolds number, one related to compressional and another to shear viscosity,
\begin{equation}
R_e^c = \frac{\rho_{*}VL}{\eta_0}, \qquad R_e^s = \frac{\rho_{*}VL}{\eta_1},
\label{eq:3.1}
\end{equation}
where $V$ is a characteristic speed, usually taken to be equal to the
phase speed of waves, $L$ a characteristic length, and $\rho_{*}$ a characteristic
density. Of course, where the viscosity is isotropic, $R_e^c = R_e^s = R_e$ and
$\vec{F}_{\rm vis}$ is given by Eq.~(\ref{eq:2.2b}). The characteristic length
$L$ plays a very important role in Eq.~(\ref{eq:3.1}). Far from resonant layers
it can be taken to be equal to the wave length. Typically in the solar corona we obtain
very large values of $R_e^c$ and $R_e^s$\/, which enables us to neglect
viscosity, for example, $R_e^s = 10^3 \div 10^4$ and 
$R_e^s = 10^{12} \div 10^{14}$.

The importance of resistivity is characterised by the magnetic Reynolds
number defined as
\begin{equation}
R_m = \frac{VL}\eta.
\label{eq:3.2}
\end{equation}
For the solar corona typically $R_m = 10^{12} \div 10^{14}$\/.

Finally, the importance of thermal conduction is characterised by the P\'eclet number which is given by 
\begin{equation}
P_e = \frac{k_B\rho_{*}VL}{m_p\kappa_\parallel}.
\label{eq:3.3}
\end{equation}
This number is very large in the solar photosphere ($10^8-10^9$), but it can be of the order 
of 100 in the solar corona.

The situation with the resistivity and thermal conduction is similar to that
of viscosity. Usually we can neglect them far from dissipative layers and
consider plasmas as infinitely conducting ($\sigma \to \infty)$ and their 
motions as adiabatic. However, these two dissipative processes can be again very
important inside the dissipative layer.

Linear theory predicts the existence of slow and Alfv\'en dissipative layers
\cite[e.g.][]{goo10}. When the dominant dissipative processes are the
isotropic plasma viscosity and resistivity, as in the solar photosphere, it is
convenient to introduce the total isotropic Reynolds number, $R_i$, defined by 
\begin{equation}
\frac1{R_i} = \frac1{R_e} + \frac1{R_m}.
\label{eq:3.4}
\end{equation}
Let us also introduce the dimensionless wave amplitude far from a dissipative layer,
$\epsilon \ll 1$. Then linear theory predicts that the amplitudes of 
perturbations of certain variables called `large variables' in a slow dissipative 
layer embracing the slow resonant position are of the order of 
$\epsilon R_i^{1/3}$\/. The linear theory can be used to describe the motion 
in the slow dissipative layer only when dissipation in this layer strongly
dominates over nonlinearity. Estimates show that the ratio of the largest
typical nonlinear term in the MHD equations to the largest dissipative term is of the
order of (see, e.g. Ruderman et al. 1997a)
\begin{equation}
N_i = \epsilon R_i^{2/3}.
\label{eq:3.5}
\end{equation}
The linear description is only valid when $N_i \ll 1$. In the solar photosphere, however, 
$\epsilon = 10^{-2} \div 10^{-3}$ and $R_i \gtrsim 10^6$\/, so that 
$N_i \gtrsim 10$. This estimate clearly shows that the motion in slow
dissipative layers is strongly nonlinear in the solar photosphere.

In the solar corona the dominant dissipative processes {\bf affecting slow waves} are the compressional
viscosity and thermal conduction (recall that it is strongly anisotropic and the
heat flux is parallel to the magnetic field). To characterise both dissipative
processes simultaneously we introduce the total anisotropic Reynolds number 
defined by (see, e.g. Ballai et al. 1998a)
\begin{equation}
\frac1{R_a} = \frac1{R_e^c} + \frac1{P_e}.
\label{eq:3.6}
\end{equation}
The linear theory predicts that the amplitude of large variables in
dissipative layers is of the order of $\epsilon R_a$\/, while the ratio of 
the largest nonlinear term in the MHD equations to the largest dissipative 
term is of the order of
\begin{equation}
N_a = \epsilon R_a^2.
\label{eq:3.7}
\end{equation}
Again, the linear description is only valid when $N_a \ll 1$. In the solar corona for a typical 
$\epsilon = 10^{-2} \div 10^{-3}$ and $R_a \gtrsim 100$\/, so we obtain 
$N_a \gtrsim 10$, and the motion in slow dissipative layers in the solar corona 
is strongly nonlinear. 

Viscosity and resistivity are also the dominant dissipative processes in 
Alfv\'en dissipative layer in the solar photosphere. Hence their importance
can be characterized by $R_i$\/. Since the quasi-Alfv\'enic motion in
Alfv\'en dissipative layers practically do not perturb the plasma temperature,
thermal conduction is not important in these layers. Since, in addition, compressional viscosity also does not operate in Alfv\'en dissipative layers,
the dominant dissipative processes in these layers in plasmas with strongly
anisotropic viscosity like in the solar corona are the shear viscosity and
resistivity. These two processes can be simultaneously described by the
total Alfv\'en Reynolds number $R_A$ defined by an equation similar to  
Eq.~(\ref{eq:3.4}) but with $R_e^s$ substituted for $R_e$\/. Applying the same
formal procedure as that resulted in Eq.~(\ref{eq:3.5}) we obtain a similar
estimate for the ratio of the typical largest nonlinear term in the MHD equations to 
the largest dissipative term $N_A$\/. On the basis of this estimate we
conclude that the linear description of Alfv\'en dissipative layers breaks down
as soon as $\epsilon \gtrsim R_A^{-2/3}$\/. However, as it will be explain
later, this conclusion is wrong as in Alfv\'en dissipative 
layers the largest nonlinear terms in the MHD equations cancel each other.

\section{Governing equation for motion in slow dissipative layers}
\label{sec:4}

In what follows we adopt Cartesian coordinates $x,\,y,\,z$ and assume that the
equilibrium quantities depend on $x$ only, while the equilibrium magnetic field
is unidirectional and parallel to the $yz$\/-plane. In the equilibrium plasma is
at rest. We only consider a two-dimensional problem and assume that perturbations
of all quantities are independent of $y$\/.  

In order to derive the equation governing the wave motion inside slow dissipative layers we
need to use the matching conditions with the solution of linear ideal MHD
equations outside the dissipative layer, which we will call the external
solution. Hence, before we embark on the derivation of the governing equation
for slow dissipative layer we briefly discuss this external solutions.

To obtain the external solution we write all variables in the system of MHD equations as a sum of an
equilibrium quantity and their Eulerian perturbation
\begin{equation}
p = p_0 + p', \quad \rho = \rho_0 + \rho', \quad \vec{B} = \vec{B}_0 + \vec{b}.
\label{eq:4.1}
\end{equation}
We substitute these expressions in the ideal MHD equations and linearize them
with respect to perturbations. We Fourier-analyse the obtained linear
equations with respect to $z$ and $t$ and take all quantities proportional to
$\exp(ikz - i\omega t)$. Finally, we eliminate all the variables from these
equations in favour of the $x$\/-component of velocity, $u$, and the total pressure
perturbation $P = p' + \vec{B}_0\cdot\vec{b}/\mu_0$ to obtain the system of two
first-order ordinary differential equations of the form
\begin{equation}
\frac{du}{dx} = \frac{i\omega P}{F}, \qquad 
\frac{dP}{dx} = \frac{ik\rho_0 D_A u}{V}.
\label{eq:4.2}
\end{equation}
In these equations $V = \omega/k$ is the phase velocity and
\begin{equation}
F = \frac{\rho_0 D_A D_C}{V^4 - V^2(v_A^2+c_S^2) + v_A^2 c_S^2\cos^2\alpha},
\label{eq:4.3}
\end{equation}
\begin{equation}
D_A = V^2 - v_A^2\cos^2\alpha, \quad
D_C = (v_A^2 + c_S^2)(V^2 - c_T^2\cos^2\alpha), 
\label{eq:4.4}
\end{equation}
where the Alfv\'en $v_A$\/, the sound $c_S$\/,  and cusp $c_T$\/, speeds are 
defined as
\begin{equation}
v_A^2 = \frac{B_0^2}{\mu_0\rho_0}, \quad c_S^2 = \frac{\gamma p_0}{\rho_0},
\quad c_T^2 = \frac{v_A^2 c_S^2}{v_A^2 + c_S^2},
\label{eq:4.5}
\end{equation}
with $\alpha$ being the angle between $\vec{B}_0$ and the $z$\/-axis, i.e.
$\cos\alpha = B_{0z}/B_0$\/.

The system of equations (\ref{eq:4.2}) has two regular singular points, $x_A$
and $x_c$\/. The first point is defined by the equation $D_A(x_A) = 0$ and
corresponds to the Alfv\'en resonance. The second point is defined by the 
equation $D_C(x_c) = 0$ and corresponds to the slow or cusp resonance. Using the
Fr\"obenius expansions one can show that the asymptotic behaviour of solution in
the vicinity of these points is given by $P \sim \mbox{const}$ and
$u \sim \mbox{const}\times\ln|x - x_{A,c}|$. Hence, $P$ is a regular function 
in the vicinity of the Alfv\'en and slow singularity. The asymptotic behaviour 
of other quantities can be determined from the linearized MHD equations.
Hence, in the vicinity of the Alfv\'en resonance, this
behaviour is given by (see, e.g. Sakurai et al. 1991; Ofman et al. 1994; Erd\'elyi and Goossens 1995; Erd\'elyi 1997, 1998)
\begin{equation}
\begin{array}{lll}
\displaystyle u \sim \mbox{const}\times\ln|x - x_A|, \quad & 
v_\parallel \sim \mbox{const}, \quad & \displaystyle 
v_\perp \sim \frac{\mbox{const}}{x - x_A}, \vspace*{1mm}\\
\displaystyle b_x \sim \mbox{const}\times\ln|x - x_A|, & 
b_\parallel \sim \mbox{const}, & \displaystyle  
b_\perp \sim \frac{\mbox{const}}{x - x_A}, \vspace*{1mm}\\
\rho' \sim \mbox{const}, & p' \sim \mbox{const}, & P \sim \mbox{const},
\end{array}
\label{eq:4.6}
\end{equation}
where $v_\parallel$ is the velocity component parallel to the equilibrium
magnetic field, and $v_\perp$ is the velocity component perpendicular both to 
the equilibrium magnetic field and to the $x$\/-direction. These two quantities are
defined by
\begin{equation}
v_\parallel = \vec{v}\cdot\vec{B}_0/B_0, \qquad
v_\perp = \vec{v}\cdot(\vec{B}_0\times\vec{e}_x)/B_0 ,
\label{eq:4.7}
\end{equation}
where $\vec{e}_x$ is the unit vector in the $x$\/-direction. The quantities 
$b_\parallel$ and $b_\perp$ can be determined in a similar way. The asymptotic behaviour of perturbations at  slow resonance are (see, e.g. 
Sakurai et al. 1991; Ballai et al. 2000a)
\begin{equation}
\begin{array}{lll}
\displaystyle u \sim \mbox{const}\times\ln|x - x_c|, \quad & \displaystyle 
v_\parallel \sim \frac{\mbox{const}}{x - x_c}, \quad & 
v_\perp \sim \mbox{const}, \vspace*{1mm}\\
\displaystyle b_x \sim \mbox{const}\times\ln|x - x_c|, & \displaystyle
b_\parallel \sim \frac{\mbox{const}}{x - x_c}, &  
b_\perp \sim \mbox{const}, \vspace*{1mm}\\
\displaystyle \rho' \sim \frac{\mbox{const}}{x - x_c}, & \displaystyle 
p' \sim \frac{\mbox{const}}{x - x_c}, & P \sim \mbox{const}.
\end{array}
\label{eq:4.8}
\end{equation}

The obvious result of Eq.~(\ref{eq:4.6}) is that the most singular variables at 
$x = x_A$ are $v_\perp$ and $b_\perp$\/ and here they have a $1/x$ singularity. Due to their behaviour at the resonant position they are called the \textit{large
variables}. In accordance with Eq.~(\ref{eq:4.8}), the large variables at 
the slow resonant position $x_c$ are $v_\parallel$\/, $b_\parallel$\/, $\rho'$
and $p'$\/.  

An important consequence of Eqs.~(\ref{eq:4.6}) and (\ref{eq:4.8}) is that the
perturbation of the total pressure, $P$\/, does not change across the
dissipative layer. The physical reason for this behaviour resides in the fact that since the dissipative
layer is thin, i.e. its inertia is very small, the acceleration of the
dissipative layer is finite, the total force applied to it has to be small. This
implies that the total pressure at the two sides of the dissipative layer has to
be almost the same. If we denote the thickness of the dissipative layer as
$\ell$ and the characteristic length of the problem as $L$\/, then the
variation of the total pressure across the dissipative layer is zero in the
zero-order approximation with respect to the small parameter $\ell/L$\/. The
conclusion that there is no jump of total pressure perturbation across the
dissipative layer is obtained here on the basis of ideal MHD. This conclusion is
confirmed by the linear theory of dissipative layers based on dissipative MHD
(e.g. Goossens et al. 1995 for Alfv\'en resonance and Erd\'elyi 1997 for slow resonance).

The linear theory suggests the following physical picture of wave motion in 
slow dissipative layers. The global motion of the plasma is in exact resonance 
with slow waves at the ideal resonant position, and it is in quasi-resonance 
with slow waves in a narrow dissipative layer embracing the resonant position. 
The variation of the external total pressure acts as a driver of slow waves in
the dissipative layer. Hence, the aim of nonlinear theory is to derive the
equation governing the nonlinear evolution of slow waves in the dissipative
layer. 

\cite{ruderman97a} were the first who considered this problem. In their
analysis they assumed that the only dissipative processes operating in the
plasma are resistivity and isotropic viscosity, and both dissipative
processes are weak, i.e. dissipation is characterized by $R_i \gg 1$. They
considered the motion in the form of a propagating wave with permanent shape, 
so that perturbations of all variables depend on the linear combination $\theta = z - Vt$\/ rather than $z$ and $t$ separately. The phase speed of
slow waves at $x_c$ is equal $c_{Tc}\cos\alpha$\/, where the subscript `$c$'
indicates that a quantity is calculated at $x = x_c$\/, so
$V = c_{Tc}\cos\alpha$\/. Finally, \cite{ruderman97a} assumed that 
perturbations of all variables are periodic with respect to $\theta$ with the
period (wave length) equal to $L$. Later \cite{ruderman00} generalized this derivation 
to allow slow time variation of the wave shape, so that perturbations of all 
variables are functions of $x$\/, $\theta$ and `slow' time $\tau$\/. 
In what follows we briefly outline this derivation. 

To obtain the governing equation that simultaneously describes both nonlinear 
and dissipative effects we formally take $N_i \sim 1$, so that 
$R_i \sim \epsilon^{-3/2}$\/. This assumption is formal in the sense that one can
consider $N_i \ll 1$, in which case we can neglect nonlinearity in comparison
with dissipation, and $N_i \ll 1$, in which case we can neglect dissipation in
comparison with nonlinearity. Linear theory predicts that the 
characteristic thickness of the dissipative layer is 
$LR_i^{1/3} \sim \epsilon^{1/2}$\/. This implies that it is convenient to
introduce the stretching variable $\xi = \epsilon^{-1/2}(x-x_c)$ in the
dissipative layer. It is also practical to introduce the `slow' time $\tau = \epsilon^{1/2}t$\/.

The nonlinear interaction of the wave motion in the dissipative layer with the
plasma causes the mean flow in the $yz$\/-plane with the amplitude of the 
order of $\epsilon^{1/2}$\/. Hence, it is convenient to split the velocity components in 
the mean and oscillating parts,
\begin{equation}
v_\parallel = \bar{v}_\parallel + \tilde{v}_\parallel, \quad
v_\perp = \bar{v}_\perp + \tilde{v}_\perp, \quad
\bar{v}_\parallel = \langle v_\parallel \rangle, \quad
\bar{v}_\perp = \langle v_\perp \rangle,
\label{eq:4.9}
\end{equation}
where the mean value of a period function $f(\theta)$ over the period is 
defined by
$$
\langle f(\theta) \rangle = \frac1L\int_0^L f(\theta)\,d\theta.
$$
It follows from Eq.~(\ref{eq:4.8}) that, in the dissipative layer,
$v_\parallel \sim \epsilon^{1/2}$\/, $b_\parallel \sim \epsilon^{1/2}$\/,
$\rho' \sim \epsilon^{1/2}$\/, $p' \sim \epsilon^{1/2}$\/, 
$u \sim \epsilon\ln\epsilon$\/, $b_x \sim \epsilon\ln\epsilon$\/,
$v_\perp \sim \epsilon$\/, $b_\perp \sim \epsilon$\/, and $P \sim \epsilon$\/.
Since $|\ln\epsilon|$ is much smaller than any negative power of $\epsilon$ for 
$\epsilon \ll 1$, in what follows we consider $\ln\epsilon$ as a quantity of the order of unity and we look for the solution to the dissipative MHD equations
in the form of power series expansions with respect to $\epsilon^{1/2}$\/. We
write this expansions in the form 
$f = \epsilon^{1/2}f_1 + \epsilon f_2 + \dots$ for $p'$\/, $\rho'$\/, 
$\tilde{v}_\parallel$\/, $b_\parallel$\/, $\bar{v}_\parallel$ and
$\bar{v}_\perp$\/, and in the form 
$g = \epsilon g_1 + \epsilon^{3/2}g_2 + \dots$ for $u$\/, 
$\tilde{v}_\perp$\/, $b_x$\/, $b_\perp$ and $P$\/.  
 
Substituting the power series expansions in the dissipative MHD equation we
obtain in the first order approximation a solution that recovers the results of the linear
theory. The compatibility condition for the equations of the second order
approximation gives the governing equation for the wave motion in the slow
dissipative layer,
\begin{equation}
2\frac{\partial\tilde{v}_\parallel}{\partial t} - \frac{\Delta_c(x - x_c)}V
  \frac{\partial\tilde{v}_\parallel}{\partial\theta}
+ \Lambda\tilde{v}_\parallel\frac{\partial\tilde{v}_\parallel}{\partial\theta}
- \lambda_i\frac{\partial^2\tilde{v}_{\parallel}}{\partial x^2}
= -\frac{c_{Tc}^2\cos\alpha}{\rho_{0c}v_{Ac}^2}\frac{dP}{d\theta},
\label{eq:4.10}
\end{equation}
where
\begin{equation}
\Delta_c = \left.\frac{d}{dx}(V^2 - c_T^2\cos^2\alpha)\right|_{x=x_c},
\label{eq:4.11} 
\end{equation}
\begin{equation}
\Lambda = \frac{v_{Ac}^2[(\gamma+1)v_{Ac}^2 + 3c_{Sc}^2]\cos\alpha}
   {(v_{Ac}^2 + c_{Sc}^2)^2} , \quad 
   \lambda_i = \frac{\eta_0}{\rho_{0c}} + \frac{c_{Tc}^2\eta}{v_{Ac}^2}.
\label{eq:4.12}
\end{equation}

When deriving Eq.~(\ref{eq:4.10}) we considered that 
$\tilde{v}_\parallel \approx \epsilon^{1/2}\tilde{v}_{\parallel1}$\/. In
Eq.~(\ref{eq:4.10}) $P$ is considered as a given function and it is determined by
the solution to the linear ideal MHD equations outside the dissipative layer.

As we have already mentioned, the main idea of connection formulae is to
consider the dissipative layer as a surface of discontinuity and use the
expressions for the jumps of $P$ and $u$ across the dissipative layer as the
boundary conditions for the system of equations (\ref{eq:4.2}). The jump of
total pressure is the same as in linear theory \cite[see, e.g.,][]{hollweg87,sak91,goossens95,ruderman96},
\begin{equation}
[P] = 0,
\label{eq:4.13}
\end{equation}
where, in general, the jump of a function $f$ across the dissipative layer is defined by
$$
[f] = \lim_{\xi \to \infty}\{f(\xi) - f(-\xi)\}.
$$
(recall that $\xi = \epsilon^{-1/2}(x - x_c)$). To calculate the expression of $[u]$ we use the
relation between $u$ and $\tilde{v}_{\parallel}$ obtained in the first order
approximation in the process of derivation of Eq.~(\ref{eq:4.10}), i.e.
\begin{equation}
\frac{\partial u}{\partial x} + \frac{c_{Tc}^2\cos\alpha}{v_{Ac}^2}
   \frac{\partial w}{\partial\theta} = 0 ,
\label{eq:4.14}
\end{equation}
where we have assumed $u \approx \epsilon u_1$ and
$\tilde{v}_\parallel \approx \epsilon^{1/2}\tilde{v}_{\parallel1}$\/. It
immediately follows from Eq.~(\ref{eq:4.14}) that
\begin{equation}
[u] = -\frac{c_{Tc}^2\cos\alpha}{v_{Ac}^2}{\cal P}
\int_{-\infty}^\infty\frac{\partial\tilde{v}_\parallel}{\partial\theta}\,dx. 
\label{eq:4.15}
\end{equation}
In the above equation $\cal P$\/, denotes the Cauchy principal part of the integral which is used since $\tilde{v}_\parallel \sim x^{-1}$ as 
$x \to \infty$\/, so that the integral in Eq.~(\ref{eq:4.14}) is divergent.

In linear theory the jump in the normal component of velocity, $[u]$, can be obtained explicitly in terms of $P$\/. In
contrast, in the nonlinear theory we, in general, cannot solve 
Eq.~(\ref{eq:4.10}) analytically and obtain an explicit expression for $[u]$,
which makes the connection formulae perhaps less practical. However we will see in
Sect.~\ref{sec:5} that there is one exception. 

\cite{ballai98a} extended the derivation by \cite{ruderman97a} where the
dominant dissipation processes are compressional viscosity and thermal
conduction along the magnetic field lines. They formally assumed that 
$N_a \sim 1$, so that $R_a \sim \epsilon^{-1/2}$\/. Only the motion periodic
with respect to $\theta$ has been considered. Using the same procedure as one
adopted by \cite{ruderman97a}, \cite{ballai98a} obtained that the 
governing equation for the motion in slow dissipative layers is
\begin{equation}
\frac{\Delta_c(x - x_c)}V \frac{\partial\tilde{v}_\parallel}{\partial\theta}
- \Lambda\tilde{v}_\parallel\frac{\partial\tilde{v}_\parallel}{\partial\theta}
+ \lambda_a\frac{\partial^2\tilde{v}_{\parallel}}{\partial\theta^2}
= \frac{c_{Tc}^2\cos\alpha}{\rho_{0c}v_{Ac}^2}\frac{dP}{d\theta},
\label{eq:4.16}
\end{equation}
where now
\begin{equation}
\lambda_a = \frac{\eta_0 V^2(2v_{Ac}^2+3c_{Sc}^2)^2}
   {3\rho_cv_{Ac}^2c_{Sc}^2(v_{Ac}^2+c_{Sc}^2)} +
   \frac{(\gamma-1)^2 m_p\kappa_\parallel V^2}{\gamma\rho_{0c}k_B c_{Sc}^2}.
\label{eq:4.17}
\end{equation}
The main difference between Eq.~(\ref{eq:4.17}) and Eq.~(\ref{eq:4.10}) is in the
terms describing the effect of dissipation, which is the last term on the
right-hand side of Eq.~(\ref{eq:4.17}), where it is proportional to the second
derivative with respect to $\theta$\/, while in Eq.~(\ref{eq:4.10}) the
corresponding term is proportional to the second derivative with respect to
$x$\/. Another difference is that Eq.~(\ref{eq:4.17}) does not contain the
derivative with respect to time. This is related to the fact that 
\cite{ballai98a} did not allow slow time variation of the wave shape in the
dissipative layer. The extension of the derivation to include this effect is
straightforward. 

The derivation of governing equations for wave motion in slow dissipative
layers has been extended to include the effect of steady flow
\cite[][]{ballai98}, cylindrical equilibrium \cite[][]{ballai00b} and the twist
of magnetic field lines \cite[][]{ballai02,erdelyi02}.

Recently, \cite{clack08} generalized the derivation of governing equation for the
wave motion in slow dissipative layers in order to take the Hall effect into account. 
The ratio of the second term on the right-hand side of the generalized Ohm's
equation (\ref{eq:2.3i}) describing the Hall effect to the first term
describing the plasma resistivity is of the order of $\omega_e\tau_e$\/. In the
solar corona typically $\omega_e\tau_e \gtrsim 10^6$\/, so that the Hall term
strongly dominates over the resistive term. In classical MHD the magnetic
field is frozen in the plasma when the plasma is infinitely conducting. When the
Hall term is taken into account the magnetic field is frozen in the electronic
component of an infinitely conducting plasma, while the ions can drift with
respect to the magnetic field lines. Hence, the account of the Hall term is
equivalent to the account of ion inertia. The MHD theory where the
Hall effect is taken into account is called the Hall MHD.

\cite{clack08} repeated the derivation carried out by \cite{ballai98a}
however in the framework of the Hall MHD. As a result they obtained that the dynamics of waves inside the dissipative layer is given by   
\begin{equation}
\frac{\Delta_c(x - x_c)}V \frac{\partial\tilde{v}_\parallel}{\partial\theta}
- \Lambda\tilde{v}_\parallel\frac{\partial\tilde{v}_\parallel}{\partial\theta}
+ \lambda_a\frac{\partial^2\tilde{v}_{\parallel}}{\partial\theta^2}
+ h_a\frac{\partial\tilde{v}_\parallel}{\partial x}
  \frac{\partial\tilde{v}_\parallel}{\partial\theta}
= \frac{c_{Tc}^2\cos\alpha}{\rho_{0c}v_{Ac}^2}\frac{dP}{d\theta},
\end{equation}
\label{eq:4.18}
where
\begin{equation}
h_a = \frac{\eta(\omega_e\tau_e)k^2 V\tan\alpha}{c_{Sc}^2 + v_{Ac}^2}.
\label{eq:4.19}
\end{equation}
In case when the dominant dissipative processes are the compressional viscosity
and thermal conduction the characteristic thickness of slow dissipative layers
given by the linear theory is  
\begin{equation}
\delta_c^a = \frac{kV\lambda_a}{|\Delta_c|}.
\label{eq:4.20}
\end{equation}
This result is easily obtained assuming that the first and third terms in
Eq.~(\ref{eq:4.18}) are of the same order. Using Eq.~(\ref{eq:4.20}) we obtain 
that the ratio of the second nonlinear term proportional to $h_a$ in 
Eq.~(\ref{eq:4.18}) to the first nonlinear term proportional to $\Lambda$ is 
of the order of
$$
\frac{\beta(\omega_e\tau_e)}{R_m}
   \left(\frac\beta{R_e^c} + \frac1{P_e}\right)^{-1},
$$
where the plasma-beta is defined by 
$\beta = c_{Sc}^2/v_{Ac}^2$\/. In particular, in the corona, where 
$\beta \lesssim 0.01$, $P_e \sim 100 < R_e^c$\/, $R_m = 10^{12} \div 10^{14}$ 
and $\omega_e\tau_e \sim 10^7$\/, this ratio does not exceed
$10^{-5}$\/. Probably, it can be more pronounced in the chromosphere where
$\omega_e\tau_e$ is already quite large, but $R_m$ is much smaller than in the
corona. 

\section{Explicit connection formulae}
\label{sec:5}

As we have seen in the previous subsection, in general, the first connection 
formula, which expresses the continuity of the total pressure, is given in
the explicit form no matter what is  the value of the nonlinearity parameter 
in a slow dissipative layer (see Eq.~(\ref{eq:4.13})). In contrast, we cannot write down an explicit expression for the jump of the normal 
component of velocity $[u]$ , in 
general. Instead, to find $[u]$, we first need to solve 
the governing equation for $\tilde{v}_\parallel$, then substitute 
$\tilde{v}_\parallel$ in Eq.~(\ref{eq:4.15}) and calculate the integral in this 
equation. As a result, we cannot separate solutions in the resonant layer and 
in the external regions. Instead, we have to solve the linear ideal MHD
equations in the external regions and the governing equation for 
$\tilde{v}_\parallel$ in the dissipative layer \emph{simultaneously} using the 
connection formulae Eqs.~(\ref{eq:4.13}) and (\ref{eq:4.15}) to connect the 
solutions. This makes solving any problem involving nonlinear slow resonance 
quite complicated.
 
However, there are two exceptions when we can obtain the second connection
formula in an explicit form. The first one is the linear theory. In this case
the connection formulae for the cylindrical geometry have been derived by 
\cite{erd95}. These formulae are easily translated to the planar geometry. We
do not give these formulae here because our aim is to study nonlinear effects 
in dissipative layer, but refer to Goossens et al. (2010) in this volume.

The second exceptional case is the 
limit of very strong nonlinearity, when nonlinearity in a slow dissipative 
layer strongly dominates over dissipation ($N_i \gg 1$ or $N_a \gg 1$). In this 
case the explicit expression for $[u]$ has been derived by \cite{rud00}. In 
what follows we briefly outline the derivation of this expression and the main points. Note that 
\cite{rud00} assumed that the equilibrium magnetic field is in the 
$z$\/-direction. Here we consider the general case when the equilibrium 
magnetic field has both $y$ and $z$\/-component, the angle between the 
$z$\/-axis and the equilibrium magnetic field being $\alpha$\/.

In what follows we consider that the dominant dissipative processes
are the isotropic viscosity and plasma resistivity. We assume that the motion
is periodic with respect to the coordinate $z$\/. so that the wave motion in the dissipative 
layer is described by Eq.~(\ref{eq:4.10}) with the time derivative equal to 
zero. Let us introduce the modified nonlinearity parameter
\begin{equation}
N = \epsilon^{3/2}(kx_0)^2 R_i,
\label{eq:5.1}
\end{equation}
where $x_0$ is the characteristic spatial scale of inhomogeneity, 
$k = 2\pi/L$\/, and $L$ is the period with respect to $z$\/. The condition
that nonlinearity dominates over dissipation in the dissipative layer is written as
$N \gg 1$. Let us introduce the dimensionless variables
\begin{equation}
\vartheta = k\theta, \;\; \sigma = -\frac{(x-x_c)\mbox{sign}(\Delta_c)}
{\delta_c(2N)^{1/3}}, \;\; 
q = \frac{V\Lambda\tilde{v}_\parallel}{\delta_c|\Delta_c|(2N)^{1/3}}, 
\;\; Q = \frac{2c_{Tc}^3\Lambda P\cos^2\alpha}
{\rho_{0c} v_{Ac}^2|\Delta_c|\delta_c (2N)^{2/3}},
\label{eq:5.2}
\end{equation}
where
\begin{equation}
\delta_c = \left[\frac{V}{k|\Delta_c|}\left(\frac{\eta_0}{\rho_{0c}} 
+ \frac{c_{Tc}^2}{v_{Ac}^2}\eta\right)\right]^{1/3} 
\label{eq:5.3}
\end{equation}
is the characteristic thickness of slow dissipative layer given by the linear
theory. In the new variables Eq.~(\ref{eq:4.10}) with
$\partial\tilde{v}_\parallel/\partial t = 0$ is rewritten as
\begin{equation}
2\sigma\frac{\partial q}{\partial\vartheta} + 
2q\frac{\partial q}{\partial\vartheta} - 
\frac1N\frac{\partial^2 q}{\sigma^2} = -\frac{dQ}{d\vartheta}.
\label{eq:5.4}
\end{equation}
Now $q$ is a periodic function of $\vartheta$ with the period $2\pi$\/. Since
$\langle \tilde{v}_\parallel \rangle = 0$ it follows $\langle q \rangle = 0$. 
It is easy to see that this restriction on $q$ is compatible with 
Eq.~(\ref{eq:5.4}). Let us look for the solution to Eq.~(\ref{eq:5.4})
satisfying this condition in the form of asymptotic expansion with respect to
the small parameter $1/N$\/,
\begin{equation}
q = \sum_{n=0}^\infty N^{-n} q^{(n)}.
\label{eq:5.5}
\end{equation}
Let us assume that the function $Q(\vartheta)$ (and, consequently, 
$P(\vartheta)$) takes its maximum value in the interval $[0,2\pi]$ at exactly 
one point $\vartheta_M$\/. We denote this value as $Q_M$\/ and the maximum 
value of $P$ as $P_M$\/. This assumption implies that 
$Q(\vartheta_M + 2\pi n) = Q_M$\/, and $Q(\vartheta) < Q_M$ when 
$\vartheta \neq \vartheta_M + 2\pi n$\/, where $n$ is any integer number. 
\cite{rud00} has shown that, under this condition, there is a quantity 
$\sigma_0 > 0$ so that the function $q^{(0)}$ can be written as
\begin{equation}
q^{(0)} = \left\{ \begin{array}{ll}
  -\sigma + [Q_M - Q(\vartheta)]^{1/2} ,\; 
  & \vartheta_M \leq \vartheta < \vartheta_S , \vspace*{3mm} \\ 
  -\sigma - [Q_M - Q(\vartheta)]^{1/2} ,\; 
  & \vartheta_S < \vartheta \leq 2\pi+\vartheta_M , \\
\end{array} \right.
\label{eq:5.6}
\end{equation}
for $|\sigma| < \sigma_0$\/. The quantity $\vartheta_S$ is a function of 
$\sigma$ implicitly defined by
\begin{equation}
\sigma = D(\vartheta_S) \equiv \frac{1}{2\pi} 
  \int_{\vartheta_M}^{\vartheta_S} [Q_M - Q(\vartheta)]^{1/2}\,d\vartheta
  - \frac{1}{2\pi} \int_{\vartheta_S}^{2\pi+\vartheta_M} 
  [Q_M - Q(\vartheta)]^{1/2}\,d\vartheta .
\label{eq:5.7}
\end{equation}
Equation~(\ref{eq:5.7}) determines $q^{(0)}$ for 
$\vartheta_M \leq \vartheta \leq \vartheta_M + 2\pi$\/. We use the periodicity
of $q^{(0)}$ with repsect to $\vartheta$ to define it beyond this interval.
When $|\sigma| > \sigma_0$\/, the function $q^{(0)}$ satisfies the condition
\begin{equation}
q^{(0)}(-\sigma,\vartheta) = -q^{(0)}(\sigma,\vartheta).
\label{eq:5.8}
\end{equation}

\begin{figure}[htbp]
\begin{center}
\includegraphics[width=8cm]{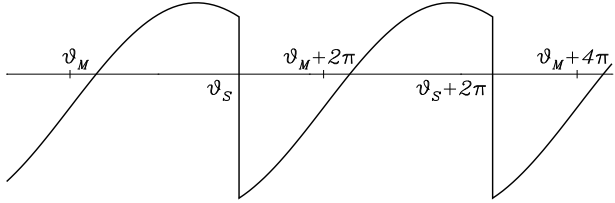}
\caption{The typical dependence of $q^{(0)}$ on $\vartheta$ at fixed $\sigma$,
$|\sigma| < \sigma_0$\/ (the figure was adapted from the study by Ruderman 2000).}
\label{fig:shock1}
\end{center}
\end{figure}

\begin{figure}
\begin{center}
\includegraphics[width=8cm]{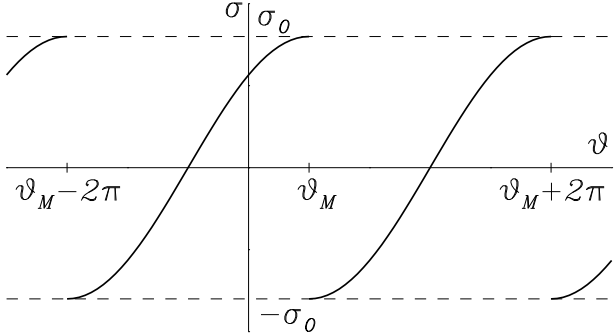}
\caption{The typical shape of slow shocks in the resonant layer (the figure was adapted from the study by Ruderman 2000).}
\label{fig:shock2}
\end{center}
\end{figure}

\cite{rud00} has shown that, in the zeroth order approximation, the wave energy
dissipation in the resonant layer occurs in slow MHD shocks
inside this layer rather than in the whole layer as it occurs in the linear
theory. The shape of shocks is defined by the equation 
$\vartheta = \vartheta_S(\sigma) + 2\pi n$\/, where $n$ is any integer number. 
The typical dependence of $q^{(0)}$ on $\vartheta$ at fixed $\sigma$,
$|\sigma| < \sigma_0$\/, is shown in Fig.~\ref{fig:shock1}. 
Figure~\ref{fig:shock2} displays the typical shape of slow shocks in the
resonant layer.
With the aid of Fig.~\ref{fig:shock2} it is easy to understand that 
Eq.~(\ref{eq:5.6}) is equivalent to
\begin{equation}
q^{(0)} = \left\{ \begin{array}{ll}
  -\sigma - [Q_M - Q(\vartheta)]^{1/2} ,\; 
  & -\sigma_0 \leq \sigma < D(\vartheta) , \vspace*{3mm} \\ 
  -\sigma + [Q_M - Q(\vartheta)]^{1/2} ,\; 
  & D(\vartheta) < \sigma \leq \sigma_0 . \\
\end{array} \right.
\label{eq:5.9}
\end{equation}
Once again, this equation defines $q^{(0)}$ for 
$\vartheta_M \leq \vartheta \leq \vartheta_M + 2\pi$\/, while $q^{(0)}$is
extended periodically beyond this interval. 

Now we are in a position to calculate $[u]$\/. Using Eq.~(\ref{eq:5.2}) we 
rewrite Eq.~(\ref{eq:4.15}) as
\begin{equation}
[u] = -\frac{kc_{Tc}\delta_c^2|\Delta_c|(2N)^{2/3}}
{v_{Ac}^2\Lambda}\frac{\partial}{\partial\vartheta}
{\cal P}\int_{-\infty}^\infty q^{(0)}\,d\sigma . 
\label{eq:5.10}
\end{equation}
We note here that, due to the relation Eq.~(\ref{eq:5.8}), the integrals of
$q^{(0)}$ over $(-\infty,-\sigma_0)$ and $(\sigma_0,\infty)$ cancel out each other. 
Then, using Eqs.~(\ref{eq:5.7}) and (\ref{eq:5.9}), we obtain
\begin{eqnarray}
{\cal P}\int_{-\infty}^\infty q^{(0)}\,d\sigma &=&
\int_{-\sigma_0}^{\sigma_0} q^{(0)}\,d\sigma
= -2D(\vartheta)[Q_M - Q(\vartheta)]^{1/2} \nonumber\\
&=& -\frac1\pi[Q_M - Q(\vartheta)]^{1/2}\bigg\{  
  \int_{\vartheta_M}^\vartheta [Q_M - Q(\tilde{\vartheta})]^{1/2}\,
  d\tilde{\vartheta} \nonumber\\ 
&-& \int_\vartheta^{2\pi+\vartheta_M} 
  [Q_M - Q(\tilde{\vartheta})]^{1/2}\,d\tilde{\vartheta}\bigg\} .
\label{eq:5.11}
\end{eqnarray}
Substituting this result into Eq.~(\ref{eq:5.10}) and returning to the original
dimensional variables, we eventually arrive at
\begin{eqnarray}
[u] &=& \frac{2kV^3 c_{Tc}^2}{\rho_{0c}v_{Ac}^4|\Delta_c|}\frac d{d\theta}
\bigg\{[P_M - P(\theta)]^{1/2} \nonumber\\
&\times& \bigg(\int_{\theta_M}^\theta 
[P_M - P(\tilde{\theta})]^{1/2}\,d\tilde{\theta} -
\int_\theta^{2\pi+\theta_M} 
  [P_M - P(\tilde{\theta})]^{1/2}\,d\tilde{\theta}\bigg)\bigg\}.
\label{eq:5.12}
\end{eqnarray}
This is the second connection formula for a strongly nonlinear slow resonant
layer. Recall that $P_M$ is the maximum value of $P$\/, $P_M = P(\theta_M)$,
and $k = 2\pi/L$\/, where $L$ is the period with respect to $\theta$\/.

Although we have used Eq.~(\ref{eq:4.10}) to derive Eq.~(\ref{eq:5.12}), the
derivation and expression for $[u]$ are exactly the same in the case where
the dominant dissipative processes are the compressional viscosity and thermal
conduction, and the plasma motion in slow dissipative layers is described by 
Eq.~(\ref{eq:4.16}). 

On the basis of analysis presented in this section we can make one very
important qualitative conclusion. The linear theory predicts that the
dimensionless amplitude of large variables in slow dissipative layer is of the
order of $\epsilon R_i^{1/3}$ in plasmas where the dominant dissipative
processes are isotropic viscosity and resistivity, and of the order of
$\epsilon R_a$ in plasmas where the dominant dissipative processes are the 
compressional viscosity and thermal conduction (recall that $\epsilon$ is the
dimensionless amplitude of wave motion far from the resonant position). Hence,
in accordance with linear theory, the amplitude of wave motion in slow 
dissipative layers tends to infinity in the limit of vanishing dissipation.

However, it follows from the analysis in this section that, in accordance with
nonlinear theory just outlined, the amplitude of wave motion in slow dissipative layers is
of the order of $\epsilon^{1/2}$ in the limit of very small dissipation. Hence,
nonlinearity causes saturation of the wave amplitude growth when the
dissipative coefficients tend to zero.

\section{Nonlinear effects in Alfv\'en dissipative layers}
\label{sec:6}

Resonant Alfv\'en waves received much more attention than their slow 
wave counterparts since for a considerable time it was thought that Alfv\'en resonance is much more important in
applications to the coronal low-beta plasma then slow resonance. In fact the hunt for finding Alfv\'en waves has 
not yet even been settled, see e.g. \cite{erd+fed07,jess09}. In the last few years the 
Alfv\'en resonance received a new connotation related to the rapid damping
of coronal loop kink oscillations, which is attributed to the resonant 
coupling of kink oscillations and local Alfv\'en waves (see the review by Andries et al. 2009).

Alfv\'en waves are incompressible and transversal, therefore the
dissipative mechanisms affecting Alfv\'en waves are the shear
viscosity and magnetic resistivity. Under coronal conditions the
coefficients describing the magnitude of these mechanisms are very
small, however dissipative terms in the momentum and induction
equation can become as large as other terms in the region containing
large spatial gradients.

Since dissipation is only important in a narrow layer embracing the 
resonant surface, the dynamics of waves {\it outside} the dissipative layer 
is described by the same system of linear MHD Eq.~(\ref{eq:4.2})
as in the case of slow waves.

The Alfv\'en resonant position, $x_A$, is defined by
\begin{equation}
V = v_A(x_A)\cos\alpha. 
\label{eq:6.1}
\end{equation}
In accordance with the linear theory the characteristic thickness of
the dissipative layer embracing the ideal resonant position is given by
\begin{equation}
\delta_A = \left(\frac{V(\eta_1+\rho_{0A}\eta)}
   {k\rho_{0A}|\Delta_A|}\right)^{1/3},
\label{eq:6.2}
\end{equation}
where now $V = v_A\cos\alpha$\/, and
\begin{equation}
\Delta_A = \left.\frac{d}{dx}(V^2 - v_A^2\cos^2\alpha)\right|_{x=x_A},
\label{eq:6.3} 
\end{equation}
and the subscript $A$ indicates that a quantity is calculated at 
$x = x_A$\/. The estimate $\delta_A \sim LR_A^{-1/3}$ immediately follows
from Eq.~(\ref{eq:6.2}), where $1/R_A = 1/R_e^s + 1/R_m$\/. In all applications
to solar physics $R_A$ is very large ranging from about $10^6$ in the
photosphere to up to $10^{14}$ in the corona.

Earlier in Sect.~\ref{sec:4} we discussed that the large variables in
Alfv\'en dissipative layers are $v_\perp$ and $b_\perp$\/. The liner theory
predicts that they are of the order of $\epsilon R_A^{-1/3}$\/, where
$\epsilon$ is the dimensionless amplitude. In the Alfv\'en dissipative layer 
the ratio of the largest nonlinear terms in the dissipative MHD equations to 
the largest dissipative terms is of the order of 
$N_A \sim \epsilon R_A^{-2/3}$\/. On the basis of this estimation we should
conclude that nonlinearity starts to compete with dissipation as soon as
$N_A \sim 1$. However, if we try to derive the governing equation for wave
motion in Alfv\'en dissipative layers using the same procedure as one adopted 
to derive Eq.~(\ref{eq:4.10}), we obtain a linear equation
\begin{equation}
(x-x_A)\frac{\partial\tilde{v}_\perp}{\partial\theta} + 
   \frac{V(\eta_1 + \rho_{0A}\eta)}{\rho_{0A}\Delta_A}
   \frac{\partial^2\tilde{v}_\perp}{\partial x^2} =
  -\frac{V\sin\alpha}{\rho_{0A}\Delta_A}\frac{dP}{d\theta},
\label{eq:6.4}
\end{equation}
where $\tilde{v}_\perp$ is the oscillatory part of $v_\perp$\/. 
The fact that Eq. (\ref{eq:6.4}) does not contain nonlinear terms despite 
considering the full nonlinear MHD system of equations is in stark 
contrast to the nonlinear description of wave motion in slow
dissipative layers presented in the previous sections. In fact,
\cite{clack09b} have shown that the linear description of wave motion in
Alfv\'en dissipative layers remains valid if $\epsilon R_A^{1/3} \ll 1$,
i.e. if the dimensionless amplitude of wave motion is much smaller than unity.
Hence, the nonlinear description is only needed when the wave motion amplitude
becomes of the order of unity. Once again note a sharp difference between the 
slow and Alfv\'en dissipative layers. In slow dissipative layer the motion 
becomes strongly nonlinear as soon as its amplitude becomes of the order of 
$\epsilon^{1/2}$\/.

The result that the linear description is valid for the wave motion in 
Alfv\'en dissipative layers is the motion amplitude is much smaller than unity
was predicted by \cite{goossens95}, but no explanation for this behaviour has 
been given. \cite{clack09a} showed that this result is related to the fact
that the largest nonlinear terms in Alfv\'en dissipative layers cancel out each
other. Using the series expansions with respect to $R_A^{-1/3}$ they also
showed that the main nonlinear effect in Alfv\'en dissipative layers is
generation of magnetosonic waves. The ratio of amplitudes of these magnetosonic
waves to the amplitude of Alfv\'enic motion is of the order of 
$\epsilon R_A^{1/3} \ll 1$.

We do not describe the results of linear theory for Alfv\'en dissipative
layers. They can be found elsewhere \cite[see e.g.][]{goo95,goo10}.

\section{Resonant interaction of externally driven waves with inhomogeneous
plasmas}
\label{sec:7}

The process of resonant absorption involves the interaction of two
oscillating systems which results in the energy transfer between
them. In this context, probably, the most obvious case is the
resonant interaction between an external laterally driven wave
and the local oscillations of the plasma. Resonant interaction between global and local
oscillations takes place if the frequency (or phase speed) of
the global wave matches any value from the local slow or Alfv\'en
continuum,  as stated in the Introduction.

The effect of nonlinearity in a slow dissipative layer on the interaction
of sound wave with an inhomogeneous magnetized plasma was first studied by
\cite{ruderman97b} and \cite{ballai98b} in the approximation of weak
nonlinearity in the dissipative layer, and by \cite{rud00} in the approximation
of strong nonlinearity. In what follows we briefly outline the analysis carried 
out in these papers.  

\begin{figure}
\begin{center}
\includegraphics[width=10cm]{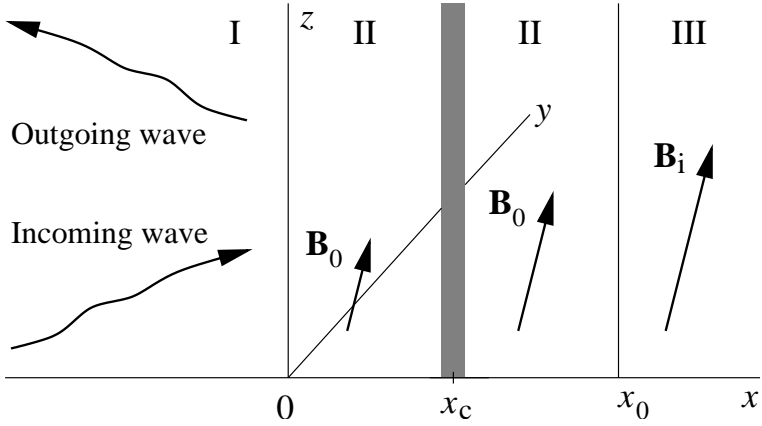}
\end{center}
\caption{The sketch of the equilibrium state. The semi-infinite regions I 
($x < 0$) and III ($x > x_0$) contain 
homogeneous plasmas penetrated by homogeneous magnetic field, while 
the plasma {and magnetic field} in region II ($0 < x < x_0$) are
inhomogeneous. The resonant surface at $x = x_c$ is inside the 
dissipative layer shown by the shaded strip (adapted from Ruderman et al. 1997b and Ballai et al. 1998b).}
\label{fig:equilib}       
\end{figure}

We consider the equilibrium schematically shown in
Fig.~\ref{fig:equilib} described with use of the Cartesian coordinates $x,y,z$\/. The 
equilibrium magnetic field is unidirectional, parallel to the
$yz$\/-plane, and makes the angle $\alpha$ with the $z$\/-axis.
The inhomogeneous plasma layer (region II) occupies the slab $0 < x < x_0$.
It is sandwiched by two semi-infinite regions containing homogeneous 
plasmas (regions~I and III), region~I being magnetic-free and region~III
being penetrated by homogeneous magnetic field. 
Obviously the model we use here is very simplistic. In reality the 
magnetic structures are much more complicated. However the approximation 
we have made will help us to understand the fundamental 
characteristics of resonant absorption and its effectiveness. The equilibrium quantities 
in region I, II and III are labelled by subscripts `$e$,' `$0$' and
`$i$', respectively.

All equilibrium quantities in region II depend on $x$ only. The 
equilibrium quantities are assumed to be continuous and satisfying the 
condition of the total pressure balance,
\begin{equation}
p_e = p_0(x) + \frac{B_0^2(x)}{2\mu_0} = p_i + \frac{B_i^2}{2\mu_0}.
\label{eq:7.1}
\end{equation}
This equation, in particular, implies that the ratio of equilibrium
densities in regions III and I is given by
\begin{equation}
\frac{\rho_i}{\rho_e} = \frac{2c_{Se}^2}{2c_{Si}^2 + \gamma v_{Ai}^2}. 
\label{eq:7.2}
\end{equation}

The plasma dynamics outside the dissipative layer is described the
system of linear ideal MHD equations. In what follows we consider
solutions in the form of propagating waves with permanent shape, and assume
that perturbations of all quantities depend on $x$ and $\theta = z - Vt$\/.
Then the system of linear ideal MHD equations can be reduces to the system of
two equations for the total pressure and $x$\/-component of the velocity,
\begin{equation}
\frac{\partial u}{\partial x} = \frac VF\frac{\partial P}{\partial\theta}, 
\qquad \frac{\partial P}{\partial x} = 
\frac{\rho_0 D_A}V\frac{\partial u}{\partial\theta},
\label{eq:7.3}
\end{equation}
where the quantities $F$ and $D_A$ are given by Eqs.~(\ref{eq:4.3}) and
(\ref{eq:4.4}). This system is written for region~II. To use it in regions~I 
and III we have to substitute $\rho_0$ by $\rho_e$ and $\rho_i$ respectively.

The physical picture of the wave interaction with the inhomogeneous plasma is
as follows. The sound wave incoming from region~I interacts with the 
inhomogeneous plasma in region~II. It is partially reflected back in region~I, 
partially penetrate in region~III, and also partially absorbed in the 
dissipative layer. As a result we have an outgoing (or reflected) wave in 
region~I in addition to the incoming wave and a transmitted wave in 
region~III.

In what follows we assume that the incoming sound wave is monochromatic.
We write its wave vector as $\vec{k} = (\chi k,0,k)$. Then its frequency is
given by
\begin{equation}
\omega^2 = k^2 c_{Se}^2(1 + \chi^2).
\label{eq:7.4}
\end{equation}
The solution to Eq.~(\ref{eq:7.3}) describing the incoming wave can be written
as
\begin{equation}
P = \epsilon p_e\cos[k(\theta + \chi x)], \quad
u = \epsilon\frac{p_e\chi}{\rho_e V}\cos[k(\theta + \chi x)],
\label{eq:7.5}
\end{equation}
where $V = \omega/k = c_{Se}(1 + \chi^2)^{1/2}$\/. In
general, the outgoing wave will contain not only the fundamental harmonic, but
also the overtones. Hence, the expression for the pressure perturbation in the
outgoing wave can be written as $P = \epsilon p_e A(\theta - \chi x)$, where 
$A$ is the function to be determined. Then the general solution to 
Eq.~(\ref{eq:7.3}) describing the wave motion in region~I is
\begin{equation}
P = \epsilon p_e\{\cos[k(\theta + \chi x)] + A(\theta - \chi x)\}, \quad
u = \epsilon\frac{p_e\chi}{\rho_e V}\{\cos[k(\theta + \chi x)]
   - A(\theta - \chi x)\}.
\label{eq:7.6}
\end{equation}

In region~III the system of Eq.~(\ref{eq:7.3}) can be reduced to
\begin{equation}
\frac{\partial P}{\partial x} + \kappa_i^2\frac{\partial P}{\partial\theta} = 0,
\quad \kappa_i^2 = -\frac{V^4 - V^2(c_{Si}^2+v_{Ai}^2) + v_{Ai}^2
c_{Si}^2\cos^2\alpha}{(c_{Si}^2 + v_{Ai}^2)(V^2 - c_{Ti}^2\cos^2\alpha)}.
\label{eq:7.7}
\end{equation}
In what follows we assume that there is a resonance position $x_c$ in
region~II, so that $V = c_T(x_c)\cos\alpha$\/. In addition we consider that
$c_T(x)$ is a monotonically increasing function in region~II. Then it is
straightforward to show that $\kappa_i^2 > 0$, so that the wave motion in 
region~III is evanescent. Using this result and expanding
$P$ in the Fourier series with respect to $\theta$ we can write the solution to
Eq.~(\ref{eq:7.3}) decaying as $x \to \infty$ in the form
\begin{equation}
P = \sum_{n=-\infty}^\infty P_n\exp[k(in\theta - \kappa_i|n|(x-x_0)],
\label{eq:7.8}
\end{equation}
where, at present, $P_n$ are arbitrary constants. Substituting this result in 
Eq.~(\ref{eq:7.3}) we obtain that the $x$\/-component of the velocity in
region~III is given by
\begin{equation}
u = \frac{iV\kappa_i}{\rho_i(V^2 - v_{Ai}^2\cos^2\alpha)}  
\sum_{n=-\infty}^\infty P_n\,{\rm sign}(n)\exp[k(in\theta-\kappa_i|n|(x-x_0))].
\label{eq:7.9}
\end{equation}

To obtain the solution in region~II we expand $P$ and $u$ in the Fourier series
with respect to $\theta$\/, and write the Fourier coefficients of these
expansions as
\begin{equation}
P_n(x) = \rho_e V^2 X_n(x) , \quad u_n(x) = iVY_n(x) .
\label{eq:7.10}
\end{equation}
Substituting these expressions in Eq.~(\ref{eq:7.3}) we obtain the system of
ordinary differential equations for $X_n$ and $Y_n$\/,
\begin{equation}
\frac{dX_n}{dx} = -\frac{\rho_0(V^2-v_A^2\cos^2\alpha)}{\rho_e V^2} nkY_n ,
\quad \frac{dY_n}{dx} = \frac{\rho_e V^2}{F} nkX_n .         
\label{eq:7.11}
\end{equation}
Consider the two solutions to this set of equations satisfying the
boundary conditions at $x=0$,
\begin{subequations}
\label{eq:7.12}
\begin{equation}
X_{1n}^{-} = 1 , \quad Y_{1n}^{-} = 0 ,
\label{eq:7.12a}
\end{equation}
\begin{equation}
X_{2n}^{-} = 0 , \quad Y_{2n}^{-} = {\rm sign}(n) ,
\label{eq:7.12b}
\end{equation}
\end{subequations}
and the two solutions satisfying the boundary conditions at $x=x_0$\/,
\begin{subequations}
\label{eq:7.13}
\begin{equation}
X_{1n}^{+} = 1 , \quad Y_{1n}^{+} = 0 ,
\label{eq:7.13a}
\end{equation}
\begin{equation}
X_{2n}^{+} = 0 , \quad Y_{2n}^{+} = {\rm sign}(n) .
\label{eq:7.13b}
\end{equation}
\end{subequations}
The solutions satisfying boundary conditions (\ref{eq:7.12}) are the two 
linearly independent solutions to the set of equations (\ref{eq:7.11}) regular 
in the interval $[0,x_c)$\/. Any other solution regular in this interval can 
be written as a liner combination of these two solutions. Similarly, the 
solutions satisfying boundary conditions (\ref{eq:7.13}) are the two linearly 
independent solutions regular in the interval $(x_c,x_0]$\/. Any other solution 
regular in this interval can be written as a liner combination of these two 
solutions.

Since the total pressure and $x$\/-component of the velocity are continuous at
$x = 0$ and $x = x_0$\/, the quantities $X_n$ and $Y_n$ satisfy the boundary
conditions at $x = 0$,
\begin{equation}
X_n = \frac{\epsilon p_e}{\rho_e V^2} \left(\frac{_1}{^2} \delta_{1|n|}
   + A_n\right) , \quad   Y_n = -\frac{i\epsilon\chi p_e}{\rho_e V^2}   
   \left(\frac{_1}{^2} \delta_{1|n|} - A_n\right) ,   
\label{eq:7.14}
\end{equation}
and at $x = x_c$\/,
\begin{equation}
X_n = \frac{P_n(x_0)}{\rho_e V^2} , \quad   
Y_n = \frac{\kappa\, {\rm sign}(n) P_n(x_0)}{\rho_i (V^2-v_{Ai}^2)} ,
\label{eq:7.15}
\end{equation}
where $\delta_{ij}$ is the Kronecker delta-symbol, and $A_n$ and $P_n$ are the
coefficients in the expansions of functions $A(\theta)$ and $P(x,\theta)$ in 
the Fourier series with respect to $\theta$\/. Using these boundary conditions 
we immediately obtain that
\begin{subequations}
\label{eq:7.16}
\begin{equation}
X_n = \frac{\epsilon p_e}{\rho_e V^2} \left[\left(\frac{_1}{^2} 
   \delta_{1|n|} + A_n\right) X_{1n}^{-} - i\chi\,{\rm sign}(n)
   \left(\frac{_1}{^2} \delta_{1|n|} - A_n\right) X_{2n}^{-}\right] ,
\label{eq:7.16a}
\end{equation}
\begin{equation}
Y_n = \frac{\epsilon p_e}{\rho_e V^2} \left[\left(\frac{_1}{^2} 
   \delta_{1|n|} + A_n\right) Y_{1n}^{-} - i\chi\,{\rm sign}(n)
   \left(\frac{_1}{^2} \delta_{1|n|} - A_n\right) Y_{2n}^{-}\right] ,
\label{eq:7.16b}
\end{equation}
\end{subequations}
in $x < x_c$\/, and
\begin{equation}
X_n = \frac{P_n(x_0)}{\rho_e V^2}\left(X_{1n}^{+} - \varsigma X_{2n}^{+}\right),
\quad
Y_n = \frac{P_n(x_0)}{\rho_e V^2}\left(Y_{1n}^{+} - \varsigma Y_{2n}^{+}\right),
\label{eq:7.17}
\end{equation}
in $x > x_c$\/, where
\begin{equation}
\varsigma = \frac{\kappa\rho_e V^2}{\rho_i (v_{Ai}^2\cos^2\alpha - V^2)} .
\label{eq:7.18}
\end{equation}
It follows from the first connection formula, Eq.~(\ref{eq:4.13}), that
$X_n(x)$ is continuous at $x = x_c$\/.

Equations (\ref{eq:7.16}) and (\ref{eq:7.17}) give the solution in region~II.
This solution contains the Fourier coefficients of the unknown function
$A(\theta)$. Now we are in a position to derive the equation for $A(\theta)$.
For this we calculate the jump of $u$ across the dissipative layer using 
Eqs.~(\ref{eq:7.16}) and (\ref{eq:7.17}), and then compare it with the jump of
$u$ given by the second connection formula. As a result we obtain the integral
equation for $A(\theta)$.

Now we need to make one final remark. The procedure of derivation of the
integral equation for $A(\theta)$ described above is only valid when the
equilibrium magnetic field is in the $z$\/-direction, i.e. when $\alpha = 0$.
The reason is the following. We have assumed that $c_T(x)$ is a monotonically
increasing function and the slow resonant position is determined by the equation
$c_T(x_c)\cos\alpha = V$\/. Since $v_A > c_T$\/, we have that $v_A(x_c)\cos\alpha > V$\/.
Since $v_A(0) = 0$, this implies that there is at leas one point 
$x_A \in (0,x_c)$ where $v_A(x_A)\cos\alpha = V$\/, i.e. there is at least one
Alfv\'en resonant position in $(0,x_c)$. There is exactly one Alfv\'en resonant
position when $v_A(x)$ is a monotonic function, while there could be a few
Alfv\'en resonant positions if $v_A(x)$ is non-monotonic. In the simplest case
when there is exactly one resonant position, $x_A < x_c$\/, we have to solve the
ideal linear MHD equations in three intervals, $(0,x_A)$, $(x_A,x_c)$ and
$(x_c,x_0)$, use the solution to calculate the jumps of $u$ at $x_A$ and 
$x_c$\/, and then compare these jumps with those given by the connection
formulae at the slow and Alfv\'en resonance. The case where $\alpha = 0$ is
exceptional because in this case there is no Alfv\'en resonance in spite that
there is $x_A$ satisfying $v_A(x_A) = V$\/.     

\subsection{Approximation of weak nonlinearity in dissipative layer}
\label{subsec:7.1}  

\cite{ruderman97b} studied the interaction of sound wave with an inhomogeneous
plasma where the dominant dissipative processes are resistivity and isotropic
viscosity under the assumption that $\alpha = 0$.  
In the nonlinear theory, in general, we cannot obtain an 
explicit expression for $[u]$\/. To make analytical progress, however, \cite{ruderman97b} 
assumed that nonlinearity in the slow dissipative layer is weak and 
considered the nonlinear term in Eq.~(\ref{eq:4.10}) as a perturbation. Then 
they carried out the asymptotic analysis using the modified nonlinearity 
parameter $N$ given by Eq.~(\ref{eq:5.1}) as a small parameter. In particular, 
they calculated the coefficient of resonant absorption defined as
\begin{equation}
K = \frac{\Pi_{\rm in} - \Pi_{\rm out}}{\Pi_{\rm in}} ,
\label{eq:7.19}
\end{equation}
where $\Pi_{\rm in}$ and $\Pi_{\rm out}$ are the energy fluxes of the 
incoming and outgoing sound waves, respectively. They found that
\begin{equation}
K \approx K_{\rm L} + N^2 K_{\rm cor},
\label{eq:7.20}
\end{equation}
where $K_{\rm L}$ is the coefficient of resonant absorption given by the linear
theory, and $N^2 K_{\rm cor}$ is the nonlinear correction, and obtained the
analytical expressions for $K_{\rm L}$ and $K_{\rm cor}$ in the approximation 
of thin inhomogeneous layer ($kx_0 \ll 1$). We do not give these expressions 
here. They can be found in \cite{ruderman97b}. We only mention that 
$K_{\rm cor} < 0$, so that nonlinearity supressess resonant absorption in the 
thin inhomogeneous layer approximation.

\cite{ballai98b} carried out a similar analysis but for plasmas where the
dominant dissipative processes are compressional viscosity and thermal
conduction, once again assuming that $\alpha = 0$. They used $(kx_0)N_a$ as a 
small parameter, where $N_a$ is given by Eq.~(\ref{eq:3.7}). \cite{ballai98b} 
obtained the expression for $K$ similar to Eq.~(\ref{eq:7.20}), however with 
$(kx_0)N_a$ substituted for $N$\/. Once again, $K_{\rm cor} < 0$ when 
$kx_0 \ll 1$, however the expression for $K_{\rm cor}$ in terms of equilibrium 
quantities is different from that obtained by \cite{ruderman97b}. The results 
obtained by \cite{ruderman97b} and \cite{ballai98b} clearly show that, while 
$K_{\rm L}$ remains the same no matter what type of dissipation operates in 
the slow dissipative layer, the coefficient of resonant absorption given by the 
nonlinear theory does depend on the dissipation type.

In a more recent study Clack and Ballai (2009a) investigated the efficiency of resonant absorption when inside the dissipative layer the 
dispersion was of the same order of magnitude as nonlinearity and dissipation. Using the limit of weak nonlinearity these authors found that the effect of dispersion is to 
increases the absolute value of the nonlinear correction (see Eq. \ref{eq:7.20}), i.e. to decreases the net coefficient of absorption. Despite the change in the nonlinear correction this term still remained small compared to 
its linear counterpart.

\subsection{Approximation of strong nonlinearity in dissipative layer}
\label{subsec:7.2}

In this subsection we review studies about the same problem of absorption of the incoming wave
energy in a slow resonant layer, however under assumption that the wave motion
in the slow dissipative layer is strongly nonlinear. To simplify the analysis we
assume that the equilibrium magnetic field is in the $z$\/-direction 
($\alpha = 0$), and region~I is magnetic-free. The latter assumption implies 
that the incoming wave is a sound wave. The detailed investigation of this 
problem is given by \cite{rud00}. Here we only outline his analysis. In what
follows we use the same notation as in \cite{rud00}.

The solution to the linear ideal MHD equations in regions~I and III, and in
region~II outside the slow dissipative layer, have been already described
above. As it has been explained, in order to obtain the equation for function
$A(\theta)$ describing the outgoing wave we have to calculate $[u]$ using the
solution of ideal linear MHD equations outside the dissipative layer, and then
compare it with the expression for $[u]$ given by the second connection
formula. In the case of strong nonlinearity in the dissipative layer this
expression is given by Eq.~(\ref{eq:5.12}). As a result we arrive at the
integral equation determining $A(\theta)$,
\begin{eqnarray}
&& M_1^r \cos(k\theta) + \chi M_1^i \sin(k\theta) + {\cal M}[A(\theta)] 
   \nonumber\\
&& \;\; = 2k\zeta [S_M - S(\theta)]^{\frac12} \left(\int_{\theta_M}^{\theta} 
   [S_M - S(\tilde{\theta})]^{\frac12} \,d\tilde{\theta} - 
   \int_{\theta}^{L+\theta_M} [S_M - S(\tilde{\theta})]^{\frac12} 
   \,d\tilde{\theta}\right) .
\label{eq:7.21}
\end{eqnarray}
Here 
\begin{equation}
S(\theta) = \frac{P_c(\theta)}{\epsilon p_e} , \quad
S_M = \frac{P_M}{\epsilon p_e} , \quad
\zeta = \frac{k\rho_e V^6}{\pi\rho_{0c}v_{Ac}^4|\Delta|} .
\label{eq:7.22}
\end{equation}
Recall that $P_M$ is the maximum value of $P$ in the dissipative layer. The
operator ${\cal M}[A(\theta)]$ is given by
\begin{equation}
{\cal M}[A(\theta)] = \sum_{n=1}^{\infty} \frac1n 
   \left[M_n A_n \exp(ink\theta) + M_n^{*} A_n^{*} \exp(-ink\theta)\right] ,
\label{eq:7.23}
\end{equation}
where $A_n$ are the Fourier coefficients of function $A(\theta)$, the asterisk
indicates the complex conjugate quantity, and the coefficients $M_1^r$\/,
$M_1^i$ and $M_n$ are expressed in terms of the equilibrium quantities and the
solutions $(X_{1n}^\pm,Y_{1n}^\pm)$ and $(X_{2n}^\pm,Y_{2n}^\pm)$ to the set of
Eq.~(\ref{eq:7.11}). We do not given these expressions here, they can be found
in \cite{rud00}. The function $S(\theta)$ is expressed in terms of $A(\theta)$ by
\begin{equation}
S(\theta) = X_{11c}^{-} \cos(k\theta) + \chi X_{21c}^{-} \sin(k\theta)
   + {\cal L}[A(\theta)] ,
\label{eq:7.24}
\end{equation}
where the operator ${\cal L}[A(\theta)]$ is given by
\begin{equation}
{\cal L}[A(\theta)] = \sum_{n=1}^{\infty} \left[(X_{1nc}^{-} +
   i\chi X_{2nc}^{-}) A_n \exp(ink\theta) + (X_{1nc}^{-} 
   - i\chi X_{2nc}^{-}) A_n^{*} \exp(-ink\theta)\right] .
\label{eq:7.25}
\end{equation}
Once again, the coefficients $X_{1nc}^{-}$ and $X_{2nc}^{-}$ are expressed in 
terms of the equilibrium quantities and the solutions to the set of
Eq.~(\ref{eq:7.11}). And, once again, we do not given these expressions here
referring to \cite{rud00} instead.

Since the operators ${\cal L}[A(\theta)]$ and ${\cal M}[A(\theta)]$ are
expressed in terms of the Fourier coefficients of the function $A(\theta)$,
and these Fourier coefficients, in turn, are expressed in terms of 
integrals of the function $A(\theta)$, the operators ${\cal L}[A(\theta)]$ and 
${\cal M}[A(\theta)]$ are integral operators. Hence, Eq.~(\ref{eq:7.21}) is 
the integral equation for the function $A(\theta)$.

In spite that Eq.~(\ref{eq:7.21}) being a very complicated nonlinear equation, it
has an extremely simple solution of the form
\begin{equation}
A(\theta) = a\cos(k\theta + \varphi) , 
\label{eq:7.26}
\end{equation}
where the quantities $a > 0$ and $\varphi$ are expressed in terms of the 
equilibrium quantities and the solutions to the set of Eq.~(\ref{eq:7.11}).
We see that the outgoing wave is monochromatic in spite that the nonlinearity in
the dissipative layer generates higher harmonics. The reason is that the
amplitudes of these higher harmonics damp at distances of the order of thickness
of the dissipative layer, so that only the fundamental harmonic survives at 
large distances from the slow resonant position.

Using the solution given by Eq.~(\ref{eq:7.26}) \cite{rud00} calculate the 
coefficient of resonant absorption. It immediately follows from 
Eq.~(\ref{eq:7.19}) that 
$K = 1 - a^2$\/. Note that this expression is valid regardless whether we use the 
linear or nonlinear description of plasma motion in the slow dissipative
layer. The nonlinearity only affects the value of $a$\/. In general, the set of
Eq.~(\ref{eq:7.11}) can be solved only numerically. However, in the thin
inhomogeneous layer approximation $(kx_0 \ll 1$) the analytical solution can be
obtained in a straightforward way. In that case the expression for the coefficient of resonant
absorption is given by
\begin{equation}
K_{\rm NL} = \frac{32\chi\zeta}{\chi^2 + \varsigma^2} + {\cal O}(k^2 x_0^2) ,
\label{eq:7.27}
\end{equation}
where we have used the subscript `NL' to indicate that the coefficient of
resonant absorption has been calculated using strongly nonlinear description of
the wave motion in the slow dissipative layer. In Eq.~(\ref{eq:7.27}) the quantity
$\varsigma$ is given by Eq.~(\ref{eq:7.18}) with $\alpha = 0$, and $\zeta$ is 
given by Eq.~(\ref{eq:7.22}). Note that $\zeta = {\cal O}(kx_0)$, so that
$K_{\rm NL} = {\cal O}(kx_0)$, which implies that resonant absorption is weak 
in the thin inhomogeneous layer approximation. For the ratio of the
coefficients of resonant absorption calculated using linear and strongly
nonlinear descriptions we obtain
\begin{equation}
\frac{K_{\rm NL}}{K_{\rm L}} = \frac{8}{\pi^2} + {\cal O}(kx_0)
   \approx 0.81 .
\label{eq:7.28}
\end{equation}
We see that, in the thin inhomogeneous layer approximation, nonlinearity
reduces the efficiency of resonant absorption. This result is in good agreement
with the results obtained in the weak nonlinearity approximation and described
in the previous subsection.

\cite{rud00} studied numerically the dependence of $K_{\rm NL}/K_{\rm L}$ on
$kx_0$ for different values of $\chi$ and a typical equilibrium. He found that 
$K_{\rm NL}/K_{\rm L}$ is a non-monotonic function of $kx_0$\/. First it grows,
takes its maximum at $k_m x_0$\/, and then monotonically decreases. Typically
$k_m x_0$ is between 4 and 6, and the maximum value of $K_{\rm NL}/K_{\rm L}$
is larger than 1. However the most important result is that, for 
$0 < kx_0 < 10$, $K_{\rm NL}/K_{\rm L}$ does not deviate from unity by mode
than 20\%. Hence, we conclude that the effect of nonlinearity in slow
dissipative layers on the coefficient of resonant absorption is moderate.

As pointed out previously, the nonlinear
coefficient of resonant absorption depends on particular dissipative processes
operating in a slow resonant layer. However, this dependence disappears in the
approximation of strong nonlinearity. This result is related to the fact that, in
the approximation of strong nonlinearity, dissipation in a slow resonant layer
occurs not in the whole volume of this layer, but in slow shocks inside the
layer. It is a very well known property of shocks that, while their internal
structure is determined by particular dissipative processes operating in a
shock, the amount of energy dissipated at the shock remains the same regardless
what the dissipative processes are. 

Recently Clack et al. (2010) introduced the concept of coupled resonances within the framework of nonlinear resonant MHD, where the transmitted waves at one of the resonances could play the role
of the incoming wave for the second resonance. It is obvious that due to the $\beta\ll 1$ in the solar corona a coupled resonance (slow+Alfv\'en) would be impossible, 
however in plasmas where the plasma-$\beta$ is much closer to unity the coupled resonance could take place. In order to ensure that the second resonance takes place before the transmitted 
waves becomes evanescent, the proximity of the two resonances must be smaller than $k^{-1}$. When looking at the efficiency of the coupled resonances Clack et al. (2010) found that 
the absorption coefficient of the coupled resonance was larger than the sum of individual coefficients taken separately at the two resonances.

\section{Generation of mean flows}
\label{sec:8}

When a wave propagates through a medium, the nonlinear interaction of the
wave and medium generates a flow. This process is very well known in nonlinear
acoustics where flows generated by propagating sound waves are called
``acoustic flows'' \cite[see e.g.][]{rud77}.

We have already mentioned about the generation of flows by resonant MHD waves in
Sect.~\ref{sec:4} when we split the velocity in the slow dissipative layer in
the mean and oscillatory parts. The flows very important for resonant and diagnostic purposes
(see e.g. \cite{doyle1997}). \cite{ruderman97a} derived the formulae 
determining the jumps of the derivative of the the mean flow across the 
dissipative layer under the assumptions that the plasma motion is periodic with
the period $L$\/, and the dominant dissipative processes are isotropic 
viscosity and resistivity. Their study was later extended by Ballai et al. (2000b) for cylindrical geometries.
These formulae determining the jumps can be written as
\begin{equation}
\left[\frac{d\bar{v}_{\perp}}{dx}\right]=
\frac{\sin\alpha}{LV}\left(1 + 
\frac{\rho_{0c}c_{Tc}^2\eta}{v_{Ac}^2\eta_0}\right)
\int_{0}^{L}d\theta\int_{-\infty}^\infty
\left(\frac{\partial\tilde{v}_\parallel}{\partial x}\right)^2 dx,
\label{eq:8.1}
\end{equation}
\begin{equation}
\left[\frac{d\bar{v}_\parallel}{dx}\right]=
-\frac{(\eta_0+\rho_{0c}\eta)V}{\eta_0 v_{Ac}^2 L\cos\alpha}
\int_{0}^{L}d\theta\int_{-\infty}^\infty
\left(\frac{\partial\tilde{v}_\parallel}{\partial x}\right)^2 dx .
\label{eq:8.2}
\end{equation}
When the motion in a slow dissipative layer can be described by the linear 
dissipative MHD equations (which is possible if $N_i \ll 1$) we can find the
explicit expression for the integrals in Eqs.~(\ref{eq:8.1}) and (\ref{eq:8.2})
in terms of perturbation of total pressure $P$\/.
In that case $\bar{v}_{\parallel}$ in the slow dissipative layer is described 
by Eq.~(\ref{eq:4.10}) with the term proportional the time derivative and the
nonlinear terms (which are the first and third term on the left-hand side) are
neglected. This linear equation can be easily solved either by the method used
by \cite{goo95} or by the method used by \cite{tirry1996} to obtain  
\begin{equation}
\tilde{v}_{\parallel n} = -\frac{ic_{Tc}V^2 P_n}{\rho_{0c}v_{Ac}^2\Delta_c}
   \left|\frac{2\pi n\Delta_c}{\lambda_i LV}\right|^{1/3}{\rm sign}(n\Delta_c)
   F\bigg((x-x_c)\left|\frac{2\pi n\Delta_c}{\lambda_i LV}\right|^{1/3}
   {\rm sign}(n\Delta_c)\bigg),
\label{eq:8.3}
\end{equation}
where $\tilde{v}_{\parallel n}$ and $P_n$ are the coefficients in the 
expansions of $\tilde{v}_\parallel$ and $P$ in the Fourier series with respect
to $\theta$\/, and the $F$\/-function is defined by
\begin{equation}
F(y) = \int_0^\infty \exp(iy\sigma - \sigma^3/3)\,d\sigma .
\label{eq:8.4}
\end{equation}
Changing the order of integration and using the Parseval identity yields
\begin{equation}
\frac1L\int_{0}^{L}d\theta\int_{-\infty}^{\infty}
\left(\frac{\partial\tilde{v}_{\parallel}}{\partial x}\right)^{2}dx
= 2\sum_{n=1}^\infty\int_{-\infty}^\infty
\left|\frac{\partial\tilde{v}_{\parallel n}}{\partial x}\right|^2 dx.
\label{eq:8.5}
\end{equation}
Using Eqs.~(\ref{eq:8.3}) and (\ref{eq:8.4}) it is not difficult to obtain that
\begin{equation}
\int_{-\infty}^\infty\left|\frac{\partial\tilde{v}_{\parallel n}}
   {\partial x}\right|^2 dx = \frac{2\pi^2 c_{Tc}^2 V^3 n|P_n|^2}
   {\rho_{0c}^2 v_{Ac}^4|\Delta_c|L\lambda_i}.
\label{eq:8.6}
\end{equation}
Substituting Eqs.~(\ref{eq:8.5}) and (\ref{eq:8.6}) in Eqs.~(\ref{eq:8.1}) and
(\ref{eq:8.2}) we arrive at
\begin{equation}
\left[\frac{d\bar{v}_{\perp}}{dx}\right] = 
   \frac{4\pi^2 c_{Tc}^2 V^2\sin\alpha}
   {\rho_{0c}v_{Ac}^4|\Delta_c|L\eta_0}
   \sum_{n=1}^\infty n|P_n|^2,
\label{eq:8.7}
\end{equation}
\begin{equation}
\left[\frac{d\bar{v}_\parallel}{dx}\right] = 
   -\frac{4\pi^2 c_{Tc}^4 V^2(\eta_0 + \rho_{0c}\eta)\cos\alpha}
   {\rho_{0c}^2 v_{Ac}^6|\Delta_c|L\eta_0}
   \left(\frac{\eta_0}{\rho_{0c}} + \frac{c_{Tc}^2\eta}{v_{Ac}^2}\right)^{-1}
   \sum_{n=1}^\infty n|P_n|^2.
\label{eq:8.8}
\end{equation}
It is instructive to estimate the order of magnitude of these jumps. It is not
difficult to show that both jumps are of the order of
$(V/L)N_i^2 R_e\,R_i^{-4/3}$\/. It is interesting that, although the amplitude 
of the wave motion in the dissipative layers is small, the jumps given by 
Eqs.~(\ref{eq:8.7}) and (\ref{eq:8.8}) can be quite large if $R_e \gg R_m$\/,
i.e. if the magnetic Prandtl number, $P_m = \eta_0/\rho_{0c}\eta$\/, is small.
Simple estimates show that these jumps are larger or of the order of $V/L$ when
$1 \ll R_m \ll R_e$ and $R_e \gtrsim \epsilon^{-2}$\/.   

To find the profiles of the components of the mean velocity we need to impose 
boundary conditions far away from the dissipative layer. For example, if there 
are rigid walls at $x = \pm a$ where the condition of adhesion has to be 
satisfied, then the components of the mean velocity take the simple form
\begin{equation}
\bar{v}_{\perp}=\left\{\begin{array}{r}
\displaystyle{\hspace*{2mm}\left[\frac{d\bar{v}_{\perp}}{dx}\right]
\frac{x-a}{2}, \quad x < 0,}\vspace*{2mm}\\
\displaystyle{-\left[\frac{d\bar{v}_{\perp}}{dx}\right]\frac{x+a}{2},
\quad x > 0,} \end{array}\right.   \qquad      
\bar{v}_{\parallel}=\left\{\begin{array}{r}
\displaystyle{\hspace*{2mm}\left[\frac{d\bar{v}_{\parallel}}{dx}\right]
\frac{x-a}{2}, \quad x < 0,}\vspace*{2mm}\\
\displaystyle{-\left[\frac{d\bar{v}_{\parallel}}{dx}\right]\frac{x+a}{2},
\quad x > 0.} \end{array}\right.      
\label{eq:8.9}    
\end{equation}
It is worth mentioning that the derivations carried in cylindrical geometry by Ballai et al. (2000b) resulted in some analogue relations for the jump in the mean flow components, but the 
properties of the mean flow generated by the nonlinear resonant slow waves are the same. This means that the results presented here are rather robust.

Although the dynamics of waves at the Alfv\'en resonance can be described within the framework of linear MHD with great accuracy, the process of resonance can produce a mean flow even in this later case. The problem of 
mean flow generation at the Alfv\'en resonance was recently studied by Clack and Ballai (2009b). In this case the jump in the oscillatory part of he parallel and perpendicular components of velocity are given as
\begin{equation}
\left[\frac{d \bar{v}_{\perp}}{dx}\right]=\frac{k^2\sin^3\alpha}{2\nu\rho_{0a}^2|\Delta_a|}\sum_{n=1}^{\infty}n|P_n|^2,
\label{eq:8.10}
\end{equation}
and
\begin{equation}
\left[\frac{d\bar{v}_{\parallel}}{dx}\right]=-\frac{k^2\sin^2\alpha\cos\alpha}{2\nu\rho_{0a}^2|\Delta_a|}\sum_{n=1}^{\infty}n|P_n|^2.
\label{eq:8.11}
\end{equation}
For the particular case of $\alpha=\pi/4$ Clack and Ballai (2009b) found that these jumps scale as $\epsilon^{1/2}V/L$. For example, if the dimensionless amplitude $\epsilon={\cal O}(10^{-4})$ then the predicted mean shear flow is of the 
order of 10 km s$^{-1}$ in both upper chromosphere and solar corona. This value is comparable with observed values of bulk flow in coronal loops. However, we should keep in mind that the important difference between these two types of flows is their range; mean shear flows are rather localised while bulk motion of plasma occurs often along the entire coronal loop length.

Studies have been carried out to investigate the properties of
shear flows, however nearly all of these have been numerical due
to analytical complications when considering nonlinearity,
turbulence and resonant absorption simultaneously. These studies
have found that shear flows could give rise to a Kelvin-Helmholtz
instability at the narrow dissipative layer \cite[see, e.g.][]{ terr08}. 
This instability can drive turbulent motions and, in
turn, locally enhance transport coefficients which can alter the
efficiency of heating \cite[see, e.g.][]{karpen94,ofman94,ofman95}. 
The generation of mean shear flow can supply additional shear enhancing 
turbulent motions.

\section{Conclusions}

Resonant absorption is a mechanism that ensures the effective transfer of energy between interacting systems. In solar and space plasmas resonant absorption is used in conjunction with the 
interaction of global waves and oscillations with inhomogeneous plasmas. Although the initial purpose of resonant absorption to be the phenomenon explaining the heating of solar corona seems these days a bit too optimistic, the advances of observational evidences of the last couple of years showed that resonant absorption is still a remarkable and brilliant mechanism on its own, able to explain a series of delicate effects occurring in the solar atmosphere and beyond.

The present review highlighted the progress made in the field of nonlinear resonance in the last 15 years. Simple estimations presented in this paper show that 
near resonances the amplitudes of oscillations can grow considerably and only a nonlinear description is adequate to describe accurately the physics near resonances. Luckily, as shown in the present review, these nonlinear additives to the effectiveness of absorption (compared to the linear counterpart) is not so significant, in general nonlinearity tends to decrease the net coefficient of absorption. In the case of resonant Alfv\'en waves, results show that the linear approach can give rather accurate results, with no need of using a mathematically more cumbersome nonlinear description.

We also showed that nonlinearity can have additional effects to the change in the absorption coefficient. Due to the nonlinear absorption of wave momentum in the vicinity of resonance, a shear mean flow is generated that is continuous across the layer containing the resonant surface, while its derivative has a jump. Simple estimations of the magnitude of the mean shear flow show that in the solar corona and upper chromosphere this mean flow is of the order of 10 km s$^{-1}$, a flow comparable with the existing bulk motion of the plasma in coronal loops. The generated shear flows can generate instabilities that can enhance locally the transport coefficients, i.e. the effectiveness of resonance. 

The analysis on the nonlinear resonant absorption reviewed by our paper contains several simplifying assumptions that helped us progress in the analytical study of the process of nonlinear resonance. It is obvious that considerable advances in this field can be made only using numerical simulations. One way to extend the existing theories is the study of nonlinear absorption in 2-D geometries. Further, all presented theories supposed (and that is true for all studies of resonant absorption, including linear cases) that the interacting waves are monochromatic, however it is more likely that waves in nature do not depend on one single wavenumber, but they depend on a spectrum of values meaning that resonant absorption would occur for all those wavenumbers for which the resonant condition is satisfied {\bf (some encouraging numerical studies were carried earlier by Ofman and Davila 1996 and Ofman et al. 1998)}. It remains to be seen what subtle effects the mean shear flow has on the resonance, probably through numerical simulations. {\bf It also known through numerical investigations (see e.g. Ofman et al. 1998) that the resonant absorption leads to the modification of the loop density structure in the nonlinear regime, thus affecting the location, dynamics, and evolution of the resonant absorption layer, as well as the heating of the loop. This aspect still needs further investigation for the cases presented in our paper. }

The abundance of high resolution observations and the extended numerical possibilities available will play an essential role in the progress of resonant absorption's study and its applicability in explaining other phenomena occurring in solar and space plasmas predicting a bright future for this simple, yet fundamental effect in inhomogeneous space and laboratory plasmas.

\acknowledgement{MSR acknowledges the support by the STFC (Science and
Technology Facilities Council) research grant. I.B. was
financially supported by NFS Hungary (OTKA, K67746) and The
National University Research Council Romania
(CNCSIS-PN-II/531/2007)
}

\bibliographystyle{aps-nameyear}
\bibliography{example}

\end{document}